\documentclass[12pt]{article}
\usepackage[nosort]{cite}
\usepackage{graphicx}
\usepackage[font={scriptsize,singlespacing}]{caption}
\usepackage{epstopdf}
\usepackage{amsmath}
\usepackage{float} %<--- allow [H] command for figure placement
\usepackage{amssymb}
\usepackage{subfigure}
\usepackage{hyperref}
\usepackage{url}
\usepackage{xcolor}

\usepackage{mathrsfs}  
\usepackage{calrsfs}

\usepackage{color}

\def\sideremark#1{\ifvmode\leavevmode\fi\vadjust{\vbox to0pt{\vss% the remark
 \hbox to 0pt{\hskip\hsize\hskip1em%                          will appear only
 \vbox{\hsize2cm\tiny\raggedright\pretolerance10000%          on the side
 \noindent #1\hfill}\hss}\vbox to8pt{\vfil}\vss}}}%
\definecolor{amaranth}{rgb}{0.9, 0.17, 0.31}
\definecolor{purple(munsell)}{rgb}{0.62, 0.0, 0.77}
\definecolor{americanrose}{rgb}{1.0, 0.01, 0.24}
\definecolor{palatinateblue}{rgb}{0.15, 0.23, 0.89}
\definecolor{royalblue(web)}{rgb}{0.25, 0.41, 0.88}
\definecolor{hanpurple}{rgb}{0.32, 0.09, 0.98}
\definecolor{beaublue}{rgb}{0.74, 0.83, 0.9}
\definecolor{carminered}{rgb}{1.0, 0.0, 0.22}
\definecolor{emerald}{rgb}{0.31, 0.78, 0.47}
\definecolor{vividviolet}{rgb}{0.62, 0.0, 1.0}
\definecolor{brightpink}{rgb}{1.0, 0.0, 0.5}
\hypersetup{ linktoc=all,
    colorlinks, linkcolor={palatinateblue},
    citecolor={brightpink}, urlcolor={amaranth}
}
\newcommand{\changeurlcolor}[1]{\hypersetup{urlcolor=#1}}

\usepackage{sectsty}
\sectionfont{\fontsize{16}{16}\bfseries\sffamily}
\subsectionfont{\fontsize{16}{16}\bfseries\sffamily}

\renewcommand{\d}[1]{\ensuremath{\operatorname{d}\!{#1}}}

\begin{document}
\thispagestyle{empty}
\begin{center}

\null \vskip-1truecm \vskip2truecm

{\LARGE{\bf \textsf{Lux in obscuro: Photon Orbits  \vskip0.25truecm of Extremal Black Holes Revisited}}}

\vskip1truecm

\textbf{\textsf{Fech Scen Khoo}}\\
\vskip0.1truecm
{Department of Physics and Earth Sciences,\\
Jacobs University, Campus Ring 1, 
28759 Bremen, Germany}\\
{\tt Email: f.khoo@jacobs-university.de}

\vskip0.6truecm

\textbf{\textsf{Yen Chin Ong}}\\
\vskip0.1truecm
{(1) Nordita, KTH Royal Institute of Technology and Stockholm University, \\ Roslagstullsbacken 23,
SE-106 91 Stockholm, Sweden\footnote{Home institute during which the work was completed. The author has moved to Shanghai Jiao Tong University, China.}}\\ \vskip0.05truecm
{(2) Riemann Center for Geometry and Physics, \\ Leibniz Universit\"at Hannover,\\
Appelstrasse 2, 30167 Hannover, Germany}\\
{\tt Email: yenchin.ong@nordita.org}\\

\end{center}
\vskip1truecm \centerline{\textsf{ABSTRACT}} \baselineskip=15pt

\medskip

It has been shown in the literature that the event horizon of an extremal asymptotically flat  Reissner-Nordstr\"om black hole is also a stable photon sphere. We further clarify this statement and give a general proof that this holds for a large class of static spherically symmetric black hole spacetimes with an extremal horizon. In contrast, in the Doran frame, an extremal asymptotically flat  Kerr black hole has an unstable photon orbit on the equatorial plane of its horizon. In addition, we show that an extremal asymptotically flat  Kerr-Newman black hole exhibits two equatorial photon orbits if $a < M/2$, one of which is on the extremal horizon in the Doran frame and is stable, whereas the second one outside the horizon is unstable. For $a > M/2$, there is only one equatorial photon orbit, located on the extremal horizon, and it is unstable. There can be no photon orbit on the horizon of a non-extremal Kerr-Newman black hole.

\section{ Black Holes and Their Photon Orbits}

One well-known feature of black holes is that due to the spacetime curvature around them, the paths of light rays can be so distorted that they end up orbiting the black holes. Such photon orbits, however, are generally unstable. That is, with a slight perturbation, light will either escape to null infinity or fall into the black hole. A textbook example is the photon orbit of an asymptotically flat Schwarzschild black hole of mass $M$, which occurs at $r=3M$ in the usual coordinate system $(t,r,\theta,\phi)$ with units $G=c=1$. (See, e.g., page 143 of \cite{wald}, or page 42 of \cite{raine}.) Due to the spherical symmetry, $r=3M$ is of course not a single orbit but a collection of infinitely many such orbits (``photon sphere''). 

If we perturb the black hole so that it starts to rotate, then there are two photon orbits on the equatorial plane, that spread out from $r=3M$. The boundaries of this region are two photon surfaces $r_1 \leqslant 3M \leqslant r_2$. (Of course the rotating case is only axisymmetric, not spherically symmetric.) Specifically \cite{BPT},
\begin{equation}\label{1}
\begin{cases}
r_1 = 2M\left[1+ \cos\left(\dfrac{2}{3}\cos^{-1}\left(-\dfrac{|a|}{M}\right)\right)\right],\\
\\
r_2 = 2M\left[1+ \cos\left(\dfrac{2}{3}\cos^{-1}\left(+\dfrac{|a|}{M}\right)\right)\right];
\end{cases}
\end{equation}
where $aM$ is the angular momentum of the black hole. The orbit at $r_1$ is prograde (moving in the same direction as the
black hole's rotation), whereas the one at $r_2$ is retrograde (moving against the black hole's
rotation). 
The motion of light rays on the photon orbits in the Kerr spacetime is, in general, rather complicated. In particular, a photon orbit need not be circular\footnote{Circular means having a constant value of $r$ in the Boyer-Lindquist coordinate. Such circular orbit is topologically $S^1$ but not geometrically a circle.}\cite{1210.2486, teo}. Henceforth, without loss of generality, we will assume that $a > 0$.

Despite their instability, photon orbits do play some important roles in black hole physics. 
It is rather surprising that even for the asymptotically flat Schwarzschild spacetime, some properties involving the photon orbit were still being discovered in the 1970s, almost 60 years after Schwarzschild's discovery. Specifically, it was found that \cite{AL,AP}, while the required outward thrust of a rocket orbiting the black hole at some constant $r> 3M$ decreases as the orbital speed of the rocket increases (in agreement with what one would expect based on Newtonian intuition), this was not the case if the orbit is below $r=3M$. For such an orbit, a \emph{greater} outward thrust is required to maintain its orbit as the orbital speed increases. (See \cite{Rickard} and the references therein for further discussion, and \cite{AP,BBI} for the generalization to stationary axisymmetric case.)

For completeness, let us give some more examples of the importance of the photon orbit. 
Firstly, the photon orbit is related to the absorption cross section of the black hole. For an asymptotically flat Schwarzschild black hole\footnote{This photon sphere is unique. See \cite{1406.5475,1504.05804}.}, the area of this geometric optics cross section is $bM^2=27\pi M^2$ \cite{wald}, where $b=\sqrt{27}M$ is the critical impact parameter for a massless particle to end up on the photon sphere.
This means that there will be a dark region that will appear in astrophysical observations of black holes, provided we have enough resolution \cite{takahashi}. This so-called ``black hole shadow'' will therefore provide tentative evidence for the existence of black holes. In fact, the Event Horizon Telescope aims to achieve precisely this \cite{EHT1, EHT2}. It is therefore crucial to understand the properties of the photon orbits, and whether non-black hole configurations can also admit such orbits. For example, it was recently shown that spherical polytropic stars cannot ``mimic'' a black hole in this way \cite{1503.01840}. In addition, observational signatures may also arise from the unusual thermodynamic properties of the photon spheres, once quantum mechanics is taken into account\cite{1410.1894}. Once quantum effects are considered, there will of course be Hawking radiation from the black hole as well. In the regime in which geometrical optics approximation is good, the effective area that emits Hawking radiation is, for an asymptotically flat Schwarzschild black hole, precisely $A=4\pi b^2$. As a consequence, the spectrum of Hawking radiation receives a greybody factor \cite{page}. 
More recently, following the direct detection of the gravitational wave by LIGO \cite{1602.03837}, it has also been emphasized that the ringdown signatures are related to the existence of photon orbits around a black hole \cite{1602.07309}. To be more specific, the frequency and damping time of the ringdown wave forms are associated with the orbital frequency and the instability timescale of the photon orbits, respectively.

Astrophysical black holes are well approximated by the asymptotically flat Kerr solution. In fact many of them are rotating so rapidly that they come extremely close to the extremal limit imposed by general relativity \cite{extreme, extreme1, extreme2}. Near-extremal black holes are therefore objects of great interest. However, it turns out that an \emph{exactly} extremal black hole behaves quite differently from a near-extremal black hole. For example, an (exactly) extremal black hole suffers from the --- classical --- Aretakis instability but near-extremal ones do not \cite{1110.2006, 1110.2007, 1110.2009, 1206.6598, 1208.1437, 1211.6903, 1307.6800}. Furthermore, the procedure of taking the extremal limit is quite subtle \cite{carroll, ingemar, stotyn}. Therefore, one expects that the properties of photon orbits in the non-extremal case (no matter how close to extremality) may differ from that of an extremal black hole case. This is exactly what one finds. Some special cases had been studied in the literature, e.g., extremal asymptotically flat Reissner--Nordstr\"om \cite{1001.0359} and Kerr black holes \cite{1108.2333}, extremal Kerr-de Sitter black hole \cite{podolsky}, as well as a type of extremal charged stringy black hole \cite{1210.0221}. There is of course a vast literature on the photon orbits of black hole spacetimes in general, see, e.g. \cite{0803.2539, 0803.2685, 0307049}. 

Such results are, however, rather perplexing. To see why, let us consider the familiar asymptotically flat Schwarzschild black hole. Any light ray that grazes the event horizon tangentially will \emph{not} stay on the surface, but instead plunges into the black hole. In other words, the event horizon is \emph{not} a photon sphere. In fact, the photon sphere is at $r=3M$, and the cylinder $\left\{r=3M\right\} \times \Bbb{R}$ is a timelike hypersurface. The world tube of event horizon, $\left\{r=2M\right\} \times \Bbb{R}$, however, corresponds to a \emph{null} hypersurface.  The character of a horizon is therefore quite different from that of the usual notion of a photon orbit. However, 
in general, what is essential when analyzing the motion of light rays, is not the timelike nature of the associated hypersurface per se, but the geodesics equations for massless particles. In this work, we will therefore define in Sec.(2) the notion of ``null photon sphere''\footnote{Note that the latter should not be confused with the ``photon surface'' defined in \cite{claudel}, since any null hypersurface, including the event horizon, would trivially be a photon surface defined therein.} via the effective potential for massless particles\footnote{The analysis of particle orbit via calculations of the critical points of an effective potential is a standard technique, and can be found in most textbooks in general relativity. Such a technique can also be useful for other purposes. See, e.g., \cite{0511057} and \cite{1005.1107}.}. We will then show that a large class of spherically symmetric static extremal black hole spacetimes possess a stable null photon sphere on their horizons. (In the literature, a stable photon sphere is also known as ``anti-photon sphere'' \cite{1608.02202}).

The most important feature of a ``null photon sphere'' is that, as we shall see, they only appear in the \emph{exactly} extremal case. This can be clearly seen in the case of an asymptotically flat Reissner-Nordstr\"om black hole --- the originally unstable photon sphere of a non-extremal black hole remains \emph{outside} the black hole in the extremal limit; and remains unstable. That is, it is \emph{not} the same photon sphere that appears on the extremal horizon, which in fact turns out to be \emph{stable} \cite{1001.0359}. Previously, stable photon orbits are known to be allowed, at least formally by the mathematics, to exist \emph{inside} the inner horizons of the asymptotically flat Kerr \cite{calvani, stuchlik} and Kerr-Newman black holes \cite{calvani2}. Since one should not trust any (naively derived) features beyond the  inner Cauchy horizons due to the mass inflation instability \cite{SimRose, poisson, 1501.04598}, these features are usually considered unphysical. 

A considerable amount of confusion has arisen in the asymptotically flat Kerr case, since in the extremal limit, whether the photon orbit and innermost stable circular orbit (ISCO) coincide with the horizon or not \emph{depends on the choice of spacelike slices} \cite{1107.5081}. In particular, in the standard Boyer-Lindquist coordinates, as can be seen from the embedding diagram (see, e.g., Fig.(2) in \cite{BPT}), these orbits are distinct but projected down to the same radius in the extremal limit as the throat tends to a cylinder (the Kerr-Newman case is explored in \cite{1503.01973}). However, as shown by Jacobson \cite{1107.5081}, in the Doran frame \cite{doran}, which is the Kerr analogue of the Painlev\'e-Gullstrand
frame for the Schwarzschild geometry, these orbits do coincide on the horizon. In this work, when we refer to photon orbits on an extremal horizon, it includes the possibility that this statement should be understood in the sense of Doran frame. 
Nevertheless, it turns out that, while the asymptotically flat Kerr black hole does possess a null photon orbit, in the sense defined in Sec.(2), such orbit is unstable. (For a fair comparison with the static black holes, for rotating black hole spacetimes, we only consider the photon orbits that are restricted to the equatorial plane.) Thus, despite some formal similarities between the charged case and the rotating case, they behave quite differently as far as the extremal photon orbits are concerned. 

What happens when both angular momentum and charge are present? Recently in \cite{1503.01973}, Ulbricht and Meinel showed that the photon orbit on the equatorial plane of an extremal asymptotically flat  Kerr-Newman black hole only lies on the event horizon if the angular momentum is sufficiently large --- more specifically, $a > M/2$. This seems to be in conflict with the previous result of \cite{1001.0359} by Pradhan and Majumdar, since in that work it is shown that an extremal asymptotically flat  Reissner-Nordstr\"om black hole does have a (stable) photon orbit on its event horizon. However, an extremal asymptotically flat  Reissner-Nordstr\"om black hole is just a special case of the Kerr-Newman family, with vanishing $a$, a value which is clearly less that $M/2$. In this work, we will also resolve these seemingly contradictory results. 

In this work we will restrict our attention to 4-dimensional spacetimes; the extension to higher dimensions --- at least for the static case --- is straightforward. 
We will also only work with spacetimes in which the black hole event horizon is the outermost horizon. This means, e.g., we do not consider Schwarzschild-de Sitter black hole, which possesses a cosmological horizon outside the event horizon. The reason for this restriction is that we have in mind an external observer who could, at least theoretically, observe the behavior of photon trajectories in the spacetime. An extremal Schwarzschild-de Sitter black hole has a degenerate event horizon that coincides with the cosmological horizon, and thus no observer in the coordinate patch can actually observe the ``black hole'' from the outside. The nature of the degenerate horizon here is also somewhat different from, say, the degenerate horizon of an extremal Reissner-Nordstr\"om black hole. In the latter case, it is the inner (Cauchy) horizon and the outer (event) horizon that coincide, and both of them are horizons associated with the black hole. On the other hand, the cosmological horizon is associated with de Sitter space, and it has a different geometrical nature (it is observer dependent) from black hole horizon. In fact, although photon orbit exists exactly on the horizon of an extremal Schwarzschild-de Sitter black hole \cite{podolsky}, one can easily show that it is \emph{not} stable (we will come back to this later).

\section{The Null Photon Sphere of Extremal Static Black Holes}

Let us start with the simple case of an asymptotically flat black hole in which the metric tensor takes the form
\begin{equation}
\d s^2 = -f(r) \d t^2 + f(r)^{-1}\d r^2 + r^2 \d\Omega^2, 
\end{equation}
where $\d\Omega^2$ need not be a metric on a 2-sphere, for example it can be a flat torus $\Bbb{R}^2/\Bbb{Z}^2$. Nevertheless, we refer to any geometry in which $g_{rr}$ and $g_{tt}$ are only functions of $r$ (and $g_{tr}\equiv 0$), as being ``spherically symmetric''.

Following the usual procedure \cite{wald, raine}, we can calculate the effective potential $V$ experienced by a massless particle. It is satisfied by the equation
\begin{equation}
V(r)=\frac{J^2}{r^2}f(r),
\end{equation}
where $J$ denotes the angular momentum of the particle. 
%Define $V:=V_{\text{eff}}^2$. The motivation to work with $V$ instead of $V_{\text{eff}}^2$ will become clear in the following discussion.

The existence of a photon orbit corresponds to a stationary point on the potential, i.e., a root of the equation
\begin{equation}\label{V'}
V'(r)=\frac{J^2}{r^2}\left[f'(r) -\frac{2}{r}f(r)\right] =0, 
\end{equation} 
where prime denotes the differentiation with respect to $r$.
Whether the photon orbit is stable or not of course depends on whether $V''$ is positive or negative. The explicit expression for $V''$ is
\begin{equation}\label{V''}
V''(r)=\frac{J^2}{r^2}\left[f''(r) - \frac{4}{r}f'(r) + \frac{6}{r^2}f(r)\right].
\end{equation}
We hereby \emph{define} ``null photon orbit'' as the solution of $f(r)=0$ such that $V'(r)=0$. 

One may worry whether this is a well-defined notion --- the vector $\partial/\partial r$ for a Reissner-Nordstr\"om black hole, for example, is timelike in between the horizons. In the neighborhood of the event horizon $r_h$, for any small $\varepsilon > 0$, $\partial/\partial r$ is timelike on $(r_h - \varepsilon, r_h)$ and spacelike on $(r_h, r_h + \varepsilon)$. It  therefore does not make sense to define a potential $V(r)$ across the horizon, let alone analyze the stability of the stationary points of the potential on the horizon. This is fortunately not a concern for an extremal black hole since $\partial/\partial r$ remains spacelike on either side of the horizon.

The existence of an extremal horizon at $r=r_h$ means that $f(r_h) = f'(r_h) = 0$. 
For a static black hole, its Hawking temperature is $T=\hbar f'(r_h)/(4\pi)$, thus extremal black holes have zero Hawking temperature. This is often taken as the working definition for an extremal black hole\footnote{It should be emphasized that the temperature of a black hole is always measured with respect to some class of observers. For an asymptotically flat spacetime, the Hawking temperature is measured by an asymptotic observer at $r \to \infty$; while for an asymptotically AdS spacetime, the Hawking temperature is interpreted as the temperature seen by the dual gauge theory on the conformal boundary of AdS, on which an observer has proper time $t$. In general, the temperature of a black hole with non-trivial asymptotic geometry is tricky to be interpreted. Nevertheless, this does not affect our argument since all that is important for our purpose is that a \emph{static} extremal black hole should satisfy $(g^{rr})'(r_h)=0=g^{rr}(r_h)$, and this together with the definition of temperature in Eq.(\ref{genhawk}) below, imposed by regularity of the Euclidean geometry, implies that $T=0$ for extremal black holes. As we have mentioned, in this work we do not consider spacetimes with a cosmological horizon, but it is worth mentioning that 
there is a debate regarding the temperature of an extremal Schwarzschild-de Sitter black hole in which a \emph{cosmological} horizon coincides with the black hole horizon \cite{9709224, 0712.3315v2} (mainly due to the subtlety of taking limit in spacetimes \cite{1502.02737}).} (see, e.g., Sec.(2.1) of\cite{livrev}). This will become a useful fact for us later on. For now, let us write the metric explicitly --- using the fact that extremal horizon is the double root of $f(r)$ --- as
\begin{equation}\label{metric1}
\d s^2 =  - \left(1-\frac{r_h}{r}\right)^2 \d t^2 + \left(1-\frac{r_h}{r}\right)^{-2}\d r^2 + r^2 \d\Omega^2.
\end{equation}
This is of course just one example --- not every static extremal black hole admits such a metric.
For this simple example, however, Eq.(\ref{V'}) now gives, for nonzero $J$,
\begin{equation}\label{12}
\left[2\left(1-\frac{r_h}{r}\right)\frac{r_h}{r^2}\right] -\frac{2}{r}\left[\left(1-\frac{r_h}{r}\right)^2\right] = \frac{2}{r}\left(1-\frac{r_h}{r}\right)\left[\frac{2r_h}{r}-1\right]  = 0,
\end{equation}
so the only solution to this equation is $r=r_h$ or $r=2r_h$. That is, there are two photon orbits, one of which is on the extremal horizon. If we substitute $f(r) = (1-r_h/r)^2$ into Eq.(\ref{V''}), we find that 
\begin{equation}
V''(r)=\frac{2J^2}{r^6}\left(10r_h^2 - 12r_h r + 3r^2\right).
\end{equation}
Local minimum requires that $V''(r) \geqslant 0$.  This leads to 
\begin{equation}
r_h \leqslant r \leqslant \frac{6-\sqrt{6}}{3} r_h \approx 1.184 ~r_h, ~~~ \text{or}~~~ r \geqslant \frac{6+\sqrt{6}}{3} r_h \approx 2.816~ r_h.
\end{equation}
This means that $r=r_h$ is a local minimum, which corresponds to a stable photon orbit; while $r=2r_h$ is a local maximum, which corresponds to an unstable photon orbit. 

In a general static spacetime, $g_{tt}g_{rr}$ need not equal $-1$ \cite{gttgrr}.
Let us write the metric as
\begin{equation}\label{metricg}
\d s^2=-\gamma(r)^2 {f}(r) \d t^2 + {f}(r)^{-1}\d r^2 + r^2\d\Omega^2; ~~{g}(r):=\gamma(r)^2{f}(r).
\end{equation}
Consider the equation of motion of a test particle with mass $m$ on this background (eventually we will set $m=0$ since we are only interested in massless particles, but for the sake of completeness we include the mass term here):
\begin{equation}
g_{\mu\nu}p^\mu p^\nu + m^2 = 0,
\end{equation}
or explicitly, assuming the motion is on the equatorial plane,
\begin{equation}
-{g}(r)\left(\frac{\d t}{\d\lambda}\right)^2 + {f}(r)^{-1}\left(\frac{\d r}{\d\lambda}\right)^2 + r^2\left(\frac{\d\phi}{\d\lambda}\right)^2 + m^2 = 0.
\end{equation}
At least for asymptotically flat geometries, analogous to the familiar Schwarzschild case, we can define the energy and angular momentum by
\begin{equation}
E_\infty :={g}(r)\frac{\d t}{\d\lambda}~~\text{and}~~J:=r^2\frac{\d\phi}{\d\lambda},
\end{equation}
respectively. Here $\lambda$ is an affine parameter that parametrizes the worldline of the particle.
Then, we obtain
\begin{equation}\label{drdphi}
\left(\frac{\d r}{\d\lambda}\right)^2 = {f}(r)\left[\frac{E_\infty^2}{{g}(r)}-\left(\frac{J^2}{r^2}+m^2\right)\right].
\end{equation}
That is,
\begin{equation}
\left(\frac{\d r}{\d\phi}\right)^2=\frac{r^4}{b^2}\left[\frac{{f}(r)}{{g}(r)}\frac{E_\infty^2}{E_\infty^2 - m^2}-{f}(r)\left(\frac{b^2}{r^2} + \frac{m^2}{E_\infty^2-m^2}\right)\right],
\end{equation}
where we have defined, as usual, the impact parameter
\begin{equation}
b:=\frac{J}{\sqrt{E_\infty^2-m^2}}.
\end{equation}
For the case of massless particles, $m=0$, we have
\begin{equation}\label{rmin}
\left(\frac{\d r}{\d\phi}\right)^2=\frac{r^4}{b^2}\left[{f}(r)\left(\frac{1}{{g(r)}} - \frac{b^2}{r^2}\right)\right].
\end{equation}
Meanwhile, from Eq.(\ref{drdphi}), we see that the effective potential is\footnote{More appropriately one should change to a new coordinate system, such that instead of the areal radius $r$, we use $R$, with $\d R^2=\dfrac{g(r)}{f(r)}\d r^2$, then we can obtain --- instead of Eq.(\ref{drdphi}) --- the usual mechanics analogue $\left(\dfrac{\d R}{\d\lambda}\right)^2 = E^2_\infty - g(r) \dfrac{J^2}{r^2}$ 
for a massless particle, 
where $r$ is now a function $r(R)$. Then, the potential is more readily identified. The result would be the same.}
\begin{equation}
V(r) = g(r)\frac{J^2}{r^2}=\gamma(r)^2{f}(r)\frac{J^2}{r^2}.
\end{equation}

The condition for the existence of a photon orbit is $V'=0$. A quick calculation shows that 
\begin{equation}\label{staticv}
V'(r)=J^2\left[\frac{2\gamma(r)\gamma'(r)f(r)}{r^2} + \frac{\gamma(r)^2f'(r)}{r^2}-\frac{2\gamma(r)^2f(r)}{r^3}\right].
\end{equation}
On the other hand, imposing regularity on the Euclidean horizon after Wick rotation, one may show that the Hawking temperature of the metric in Eq.(\ref{metricg}) is 
\begin{equation}\label{genhawk}
T=\frac{\hbar}{2\pi}\left[\frac{\gamma(r)}{2}\frac{\d f}{\d r} + f(r)\frac{\d\gamma}{\d r}\right]_{r=r_h}.
\end{equation}
At extremality, the Hawking temperature vanishes, which means that
\begin{equation}\label{extrel}
\frac{\gamma(r_h)}{2}f'(r_h)=-f(r_h)\gamma'(r_h).
\end{equation}
This implies that for an extremal horizon, Eq.(\ref{staticv}) will reduce to
\begin{equation}
V'(r_h)=-\frac{2\gamma(r_h)^2f(r_h)J^2}{r_h^3}.
\end{equation}
This vanishes on the horizon since $f(r_h)=0$. That is to say, the extremal horizon is a photon sphere. The second derivative of $V$ satisfies 
\begin{equation}
V''(r_h)=\frac{J^2\gamma(r_h)^2f''(r_h)}{r_h^2}.
\end{equation}
If this is positive on the horizon, then we are done. Indeed, if the metric is given by Eq.(\ref{metric1}), then $f''(r_h)$ is evidently positive. For the general case, if we assume the metric is real analytic\footnote{Analyticity is a stronger statement than smoothness, since it implies a certain rigidity property: if two analytic mappings between two manifolds $f,g : \mathcal{M} \to \mathcal{N}$ agree on some open set $\mathcal{U} \subset \mathcal{M}$, and $\mathcal{M}$ is connected, then $f$ and $g$ are globally identical. Admittedly this seems too strong as a physical requirement, although it is one of the premises in the proof of the uniqueness of the Kerr solution (Carter-Hawking-Robinson theorem \cite{HCR1,HCR2,HCR3}). (A lot of efforts are now being put into improving the proof without assuming analyticity \cite{K1,K2}.)  A pragmatic physicist who is used to power series solutions might be less concerned with this technical assumption. We also note that static or electrovacuum spacetime can always be endowed with analytic charts if there is a timelike Killing vector field \cite{zum}. This condition, however, fails on black hole horizons. (Static vacuum spacetimes with \emph{non-degenerate} horizons are one-sided analytic \cite{0402087}.)}, then this follows from noting that the Taylor series of $f(r)$ at the extremal horizon is
\begin{equation}
f(r) = \sum_{n=2}^\infty \frac{f^{(n)}(r_h)}{n!} (r-r_{h})^n,
\end{equation}
because $f(r_h)=f'(r_h)=0$. 
First, let us suppose that $f''(r_h) \neq 0$.
Since $f(r) > 0$ for all $r>r_h$, we must have $f''(r_h) > 0$, otherwise for sufficiently small $r-r_h$, the metric coefficient $f(r)$ would become negative\footnote{This part of the argument breaks down for an extremal Schwarzschild-de Sitter black hole. This is because in de-Sitter spacetime in the static coordinates, the coordinate $r$ is bounded: $r \leqslant L$.}. 

A more precise statement can be made by using the Lagrange remainder. Specifically, 
\begin{equation}
f(r) = \sum_{n=2}^\infty \frac{f^{(n)}(r_h)}{n!} (r-r_{h})^n = \frac{(r-r_h)^2}{2}f''(r_h) + \mathcal{R},
\end{equation}
where the remainder $\mathcal{R}$ is given by
\begin{equation}
\mathcal{R} = \frac{f^{'''}(r^*)}{3!} (r-r_h)^3,
\end{equation}
where $r^* \in (r_h, r)$. 
If $r > r_h$, we have $f(r)>0$, so 
\begin{equation}\label{ineq}
f''(r_h) + \frac{f^{'''}(r^*)}{3} (r-r_h) > 0.
\end{equation}
Suppose to the contrary that $f''(r_h) = -\mathcal{C}$ for some positive $\mathcal{C}$, then
 $f^{'''}(r^*)$ has to be positive. 
We see that inequality (\ref{ineq}) is equivalent to the condition that
\begin{equation}
r-r_h > \frac{3\mathcal{C}}{f^{'''}(r^*)} =:\delta.
\end{equation}
But $r-r_h$ can be arbitrarily small; if it is smaller than $\delta$, it now follows that we cannot have $f(r)>0$ for $r>r_h$. This is a contradiction, and 
thus we must have $f''(r_h) > 0$, as claimed. 

In the unlikely case that $f''(r_h)$ also vanishes, we need to use higher derivative test. 
By a similar argument, the fist nonzero Taylor coefficient must be positive. If this coefficient $f^{(k)}(r_h)/k!$ is such that $k$ is even, then we have a local minimum and the photon orbit is stable. However if $k$ is odd, then instead of being a local minimum, we would have a strictly increasing  inflection point, and the photon orbit is therefore unstable. Thus, we obtain the following result: 
\begin{quote}
\textbf{Theorem:} \textit{Suppose a black hole spacetime is spherically symmetric, i.e., it admits a metric tensor of the form 
\begin{equation}
\d s^2=-\gamma(r)^2 {f}(r) \d t^2 + {f}(r)^{-1}\d r^2 + r^2\d\Omega^2, \notag
\end{equation}
and furthermore that the event horizon $r_h$ is the outermost horizon in the spacetime,
then there is a photon orbit on $r_h$ when the black hole is extremal. That is, 
\begin{equation}
f(r_h)=0=f'(r_h). \notag
\end{equation} 
Furthermore, assuming that $f$ is a real analytic function, the photon orbit is
\begin{itemize}
\item[(1)] stable, if the first non-vanishing Taylor coefficient $f^{(k)}(r_h)/k!$ is with an even $k$, $k \geqslant 2$; and
\item[(2)] unstable, if otherwise.
\end{itemize}}
\end{quote}

\subsection{Special Case I: Asymptotically Flat Reissner-Nordstr\"om Black Hole}\label{sss}

Now let us examine the photon sphere for an asymptotically flat Reissner-Nordstr\"om black hole, with metric tensor of the form
\begin{equation}
\d s^2 = -\left(1-\frac{2M}{r} + \frac{Q^2}{r^2}\right)\d t^2+\left(1-\frac{2M}{r} + \frac{Q^2}{r^2}\right)^{-1}\d r^2 + r^2 \left(\d\theta^2 + \sin^2\theta \d\phi^2\right).
\end{equation}
The effective potential for massless particles is 
\begin{equation}
V(r)=\frac{J^2}{r^2}\left(1-\frac{2M}{r}+\frac{Q^2}{r^2}\right).
\label{RN_V}
\end{equation}
Solving for $V'=0$, we find that the photon sphere is at
\begin{equation}\label{outerRN}
r^+_{\text{ph}} = \frac{3M + \sqrt{9M^2-8Q^2}}{2}.
\end{equation}
Note that the other root
\begin{equation}
r^-_{\text{ph}}= \frac{3M - \sqrt{9M^2-8Q^2}}{2}
\end{equation}
\emph{cannot} be a photon sphere since it lies between the two horizons. In that region of spacetime, $\partial/\partial r$ is a timelike vector and so even though $V$ has a formal local minimum there, $r=r^-_{\text{ph}}$ cannot be interpreted as an ``orbit'' (see also \cite{claudel}). 

\begin{figure}[H]
\begin{center}
\includegraphics[width=3.1in]{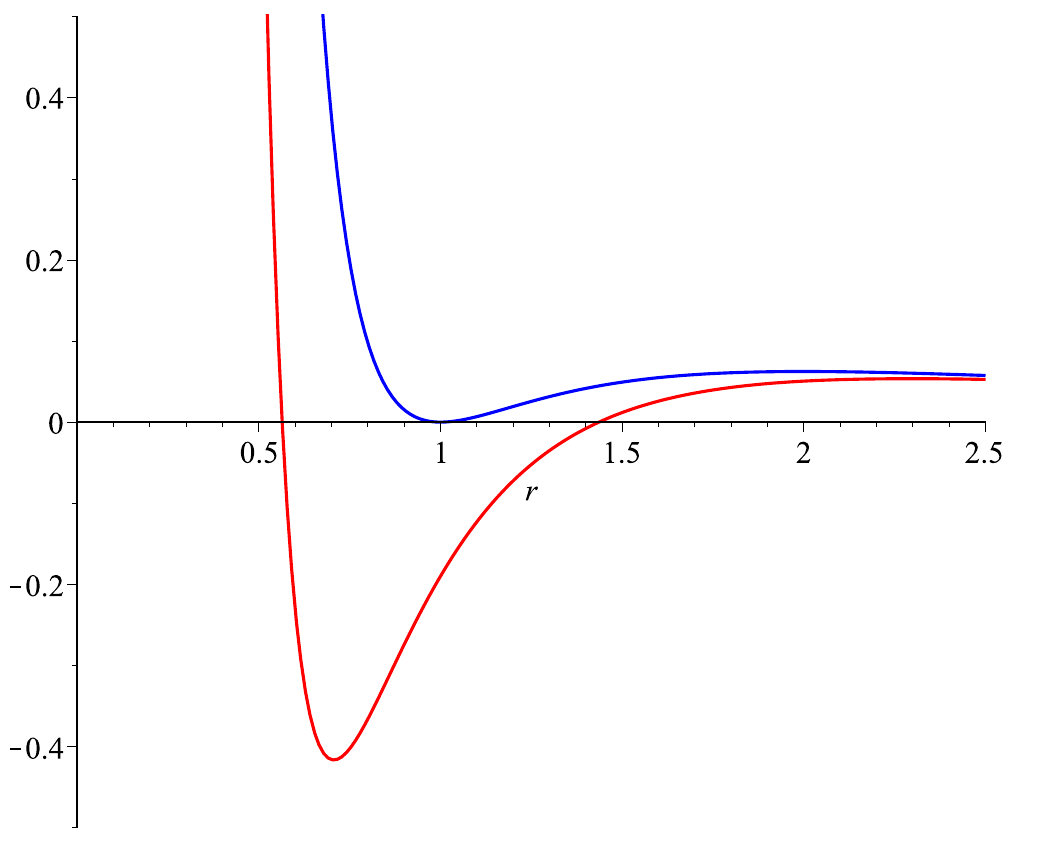} 
\caption{The plots of the effective potential of a non-extremal Reissner-Nordstr\"om black hole with spherical horizon with generic parameters (bottom curve), and that of the extremal one (top curve). The latter has a local minimum at the extremal horizon ($r_h=1$ in this example). Though not obvious in the plot, $r=2$ corresponds to a local maximum for the blue curve. In the non-extremal case, the local minimum is in between the horizons, a region in which $\partial/\partial r$ is timelike, and thus does
not correspond to a photon orbit. The (stable) photon orbit interpretation is only possible at extremality, when the local
minimum coincides with the degenerate double root of the potential. \label{RNplot}
} 
\end{center}
\end{figure}

In the extremal limit $M \to Q$, we see that $r^+_{\text{ph}} \to 2M$, which remains outside the extremal horizon $r_h = M$. 
This agrees with Eq.(\ref{12}), since the extremal asymptotically flat Reissner-Nordstr\"om black hole has a metric tensor precisely of the form given in Eq.(\ref{metric1}).
However, note that $r^-_{\text{ph}}$, which previously is not physical, now tends to $M$, which is the same position as the extremal horizon. In other words, in the exactly extremal case, 
$r^-_{\text{ph}}$ is now a physical photon orbit, which exists independently of the ``outer'' photon orbit  $r^+_{\text{ph}}$. See Fig.(\ref{RNplot}). This is the result obtained in \cite{1001.0359}.

\subsection{Special Case II: Asymptotically Locally AdS Reissner-Nordstr\"om Black Hole with Toral Horizon}

The metric of an AdS black hole with a toral event horizon is, in the Lorentz-Heaviside units, 
\begin{equation}
\d s^2  = -\, \Bigg({r^2\over L^2}\;-\;{8 \pi M^*\over r}+{4\pi Q^{*2}\over r^2}\Bigg)\d t^2\; +\Bigg( {r^2\over L^2}\;-\;{8 \pi M^*\over r}+{4\pi Q^{*2}\over r^2}\Bigg)^{-1}\d r^2+\;r^2\Big(\d\zeta^2\;+\;\d\xi^2\Big),
\end{equation}
where $\zeta, \xi \in [0, 2\pi K)$  are dimensionless coordinates on a flat 2-dimensional Riemannian manifold (and thus has zero spatial scalar curvature). $M^*$ and $Q^*$ are mass and charge parameters, respectively.  Let us take the horizon to be a flat square torus with area 4$\pi^2 K^2 r_h^2$, in which $K$ is a dimensionless ``compactification parameter'', and $r_h$ the event horizon. Then $M^* =M/(4\pi^2 K^2)$, and $Q^* = Q/(4\pi^2 K^2)$, where $M$ and $Q$ are the physical mass and charge of the black hole. 

The potential for massless particles is given by
\begin{equation}
V(r)=\frac{J^2}{r^2}\left(\frac{r^2}{L^2}-\frac{8\pi M^*}{r}+\frac{4\pi {Q^*}^2}{r^2}\right).
\end{equation}
This function is monotonically increasing outside the horizon. In other words, it does not have any local maximum.
This implies that there is no photon orbit outside the event horizon, stable or otherwise. This is due to the underlying flat spatial geometry; a spherical black hole in AdS would still have an unstable orbit. (For the uncharged case, the photon orbit of a spherical Schwarzschild-AdS black hole is at $r=3M$. See \cite{0408016v1}).

\begin{figure}[H]
\begin{center}
\includegraphics[width=3.1in]{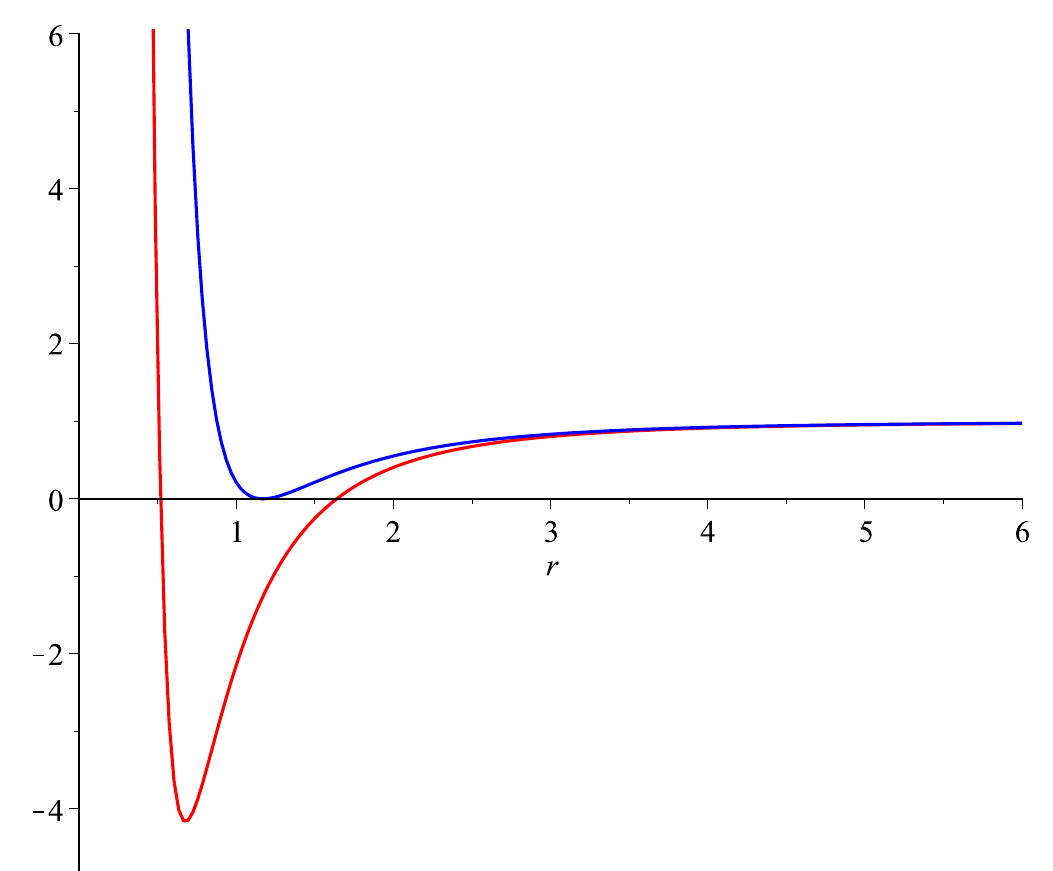} 
\caption{The plots of the effective potential of a non-extremal charged toral black hole with generic parameters (bottom curve), and that of the extremal one (top curve). In the non-extremal case, the local minimum is in between the horizons and thus does not correspond to a photon orbit. The (stable) photon orbit interpretation is only possible at extremality, when the local minimum coincides with the degenerate double root of the potential. \label{flatRNplot}} 
\end{center}
\end{figure}

It is a simple exercise to show that, despite an absence of any photon orbit in the non-extremal case, 
if the toral black hole is extremally charged, its extremal horizon admits null photon orbits in the sense we have defined in the beginning of this section.
This is illustrated in Fig.(\ref{flatRNplot}), which is a numerical plot of the effective potential.

In fact, the result also holds for other topological black holes in AdS, see appendix for the elementary, but explicit, calculation.

\section{A Non-Example: Charged Dilaton Black Hole}

In the low energy limit of string theory, a scalar field known as dilaton couples with the electromagnetic field. It admits an asymptotically flat charged black hole solution of the form \cite{GHS, GM, darkside}
\begin{equation}
\d s^2 = -\left(1-\frac{2M}{r}\right) \d t^2 + \left(1-\frac{2M}{r}\right)^{-1} \d r^2 +  r\left(r-\frac{Q^2}{M}\right) \left(\d\theta^2 + \sin^2\theta \d\phi^2\right),
\end{equation}
if the asymptotic dilaton field value is zero.
The dilaton $\varphi$ relates to the Maxwell field $F=Q \sin \theta \d\theta \wedge \d\phi$ by 
\begin{equation}
e^{-2\varphi}= 1-\frac{Q^2}{Mr}.
\end{equation}

This metric looks almost the same as that of an asymptotically flat Schwarzschild black hole. However, unlike the Schwarzschild metric, the coordinate $r$ is not a radial radius. 

The geometry we are discussing here is the so-called Einstein's frame geometry. 
In the string frame metric, which is related to the Einstein's frame metric by a conformal factor, the naked null singularity becomes a well-defined sphere. The string worldsheet has a minimal surface area with respect to the string frame metric. For a study of null geodesics in charged dilaton black hole spacetime in both the Einstein's frame and the string frame, see \cite{1109.0254}.

The effective potential for massless particles in this black hole geometry is
\begin{equation}
V(r) = \frac{J^2}{r\left(r-\frac{Q^2}{M}\right)}\left(1-\frac{2M}{r}\right).
\end{equation} 
The first derivative is
\begin{equation}
V'(r) = \frac{J^2 M[6M^2r-2M(2Q^2+r^2)+Q^2r]}{r^3(Mr-Q^2)^2}.
\end{equation} 
The photon sphere is situated at
\begin{equation}
r_{\text{ph}}^+ = \frac{\sqrt{36M^4 - 20M^2Q^2 +Q^4} + 6M^2 +Q^2}{4M}.
\end{equation}
Note that the numerator of $V'(r)$ has two zeroes $r_{\text{ph}}^\pm$, but ${ r_{\text{ph}}^-}$ is never physical, see the detailed analysis in \cite{claudel}.

Indeed we see that in the extremal limit  ($Q^2 \to 2M^2$), the photon sphere tends to the extremal horizon ($r_\text{ph}^+ \to 2M$). 
The second derivative of the effective potential is
\begin{equation}
V''(r)=-\frac{2MJ^2(12M^3r^2-16M^2Q^2r-3M^2r^3+6MQ^4+3MQ^2r^2-Q^4r)}{r^4(Mr-Q^2)^3}.
\end{equation}
This expression goes like $V'' \to -\infty$ as the extremal limit is approached. That is, naively one would say that there is an \emph{unstable} photon orbit on the extremal horizon. See also \cite{1210.0221}. 

This seems to contradict our previous result that extremal static black holes have stable null photon orbit on their event horizon. However, there is no real contradiction here since the extremal dilaton charged black hole 
is, strictly speaking, not a black hole at all --- its event horizon has degenerated into a naked null singularity, and cannot be interpreted as a photon sphere. Furthermore, our argument for the existence of a stable null photon orbit really depends on the extremal horizon being a degenerate horizon, in the sense that the inner horizon coincides with the outer, event horizon. In the case of charged dilaton black hole there is only one horizon and so the argument breaks down. (Note that, unfortunately, the word ``degenerate'' has two different meanings in this discussion.) In addition, the Hawking temperature of the extremal black hole is \emph{non-vanishing}. (See also \cite{9202014}, in which it was also pointed out that the effective potential of a massless scalar field diverges at the extremal ``horizon''). The Hawking temperature is $\hbar/8\pi M$, which is the same as a Schwarzschild black hole of the same mass. (In fact its Hawking temperature is independent of the amount of electrical charge and is \emph{always} equal to $\hbar/8\pi M$). So indeed the extremal dilaton charged ``black hole'' is not an extremal black hole in the usual sense.

We remark that in a ``mechanics'' problem such as analyzing the orbital movement, one may find it more intuitive to work with a coordinate such that the 2-sphere part of the metric takes the form $R^2\d\Omega^2$, so that $R=R(r)$ \emph{is} a radial radius. However, the result here of course does not depend on the choice of coordinates.

\section{Rotating Black Holes in General Relativity}

In the preceding section we have shown that extremal static black holes have photon orbits on their event horizon. Remarkably these photon orbits are also stable. This is a very generic statement applicable to a wide class of black hole solutions. What happens if the black hole is stationary but not static? It is well-known that spacetime around rotating black holes are much more complicated to analyze. Therefore, instead of aiming to understand a general class of rotating spacetimes, let us focus on two concrete examples: the asymptotically flat Kerr and Kerr-Newman black holes in general relativity. Since we are mainly interested in photon orbits on the horizon, and therefore well inside the ergosphere (which extends up to $r=2M$ in the equatorial plane for the extremal Kerr spacetime), we are only considering co-rotating photon orbits.

\subsection{The Case of Asymptotically Flat Kerr Black Hole}

Let us now comment on the case of asymptotically flat Kerr black holes. The metric tensor in the standard Boyer-Lindquist coordinates is
\begin{flalign}\label{kerrmetric}
\d s^2 =&-\left(1-\frac{2Mr}{\rho^2}\right)\d t^2 - \frac{4Mar\sin^2\theta}{\rho^2} \d t\d\phi + \frac{\rho^2}{\Delta^2} \d r^2 +\rho^2 \d\theta^2  \\&+\frac{\sin^2\theta}{\rho^2}\left[(r^2+a^2)^2 - a^2 \Delta \sin^2\theta\right]\d\phi^2,
\end{flalign}
where $\Delta(r):=r^2-2Mr+a^2$ and $\rho^2(r,\theta):=r^2 + a^2 \cos^2\theta$. Note that for fixed $t$ and fixed $r$, the submanifold corresponds to a topological 2-sphere, but it is not geometrically a round sphere. 

While the static spacetimes discussed thus far are spherically symmetric, a rotating black hole such as an asymptotically flat Kerr black hole is only axisymmetric. The motion of a massless particle around a rotating black hole is much more complicated, so to have a fair and direct comparison with the static cases, we shall restrict to the equatorial plane\footnote{Geodesics on the equatorial plane enjoy various nice properties. For example, its Carter constant $\mathcal{Q}$ vanishes identically. This does not completely determine the equatorial geodesics however, since many geodesics with $\mathcal{Q}=0$ are not equatorial. In fact, through every point on the Kerr spacetime, there exists a geodesic with $\mathcal{Q}=0$ that passes through it. See  Section 4.14, and Corollary 4.5.7, of \cite{O'Neill}.}.  

On the equatorial plane, the ``effective potential'' for massless particles can be derived \cite{hobson} by considering the equation
\begin{equation}\label{momenta}
g^{tt}(p_t)^2 + 2g^{t\phi}p_t p_\phi + g^{\phi\phi} (p_\phi)^2 + g^{rr}(p_r)^2 = 0.
\end{equation}
Indeed, given a spacetime $(\mathcal{M},g)$, furnished with a coordinate chart $\left\{x^\mu\right\}$, one could write down the ``Lagrangian'' $\mathcal{L}(x^\mu,\dot{x}^\mu)=g_{\mu\nu}\dot{x}^\mu \dot{x}^\nu$. Here dot denotes differentiation with respect to the affine parameter $\lambda$ that parametrizes the particle worldline. An appropriate affine parameter along the null geodesic can be chosen such that the conjugate momenta $p_\mu$ in Eq.(\ref{momenta}) satisfies $\dot{x}^\mu=g^{\mu\nu}p_\nu=p^\mu$ \cite{hobson} . The trick is to re-write Eq.(\ref{momenta}) into the standard ``mechanic form'' 
\begin{equation}
\dot{r}^2 + V(r,a,b)=E_{\infty}^2. 
\end{equation}

The effective potential turns out to be\footnote{Our definition of the effective potential differs from that of \cite{hobson} by an overall factor of $J^2$.}
\begin{equation}
V(r,a,b)=\frac{J^2}{r^2}\left[1-\left(\frac{a}{b}\right)^2-\frac{2M}{r}\left(1-\frac{a}{b}\right)^2\right],
\label{K_V}
\end{equation}
where $b:=J/E_{\infty}$ plays the role of an impact parameter of the particle with energy $E_\infty$ and angular momentum per unit rest mass $J$. As emphasized in \cite{hobson}, one must be careful in interpreting $V$ as an effective potential due to its $b$-dependence. Nevertheless, the stability of a circular photon orbit can still be studied in the usual manner by taking the second derivative of $V$ with respect to $r$.

Note that in order to have the photon orbit on the extremal horizon, the impact parameter $b$ must take a special value \cite{hobson}. This is easily seen by solving for $V'=0$ and setting the root to $r_h=M$, which yields $b=2M$.

\begin{figure}[H]
\begin{center}
\includegraphics[width=3.0in]{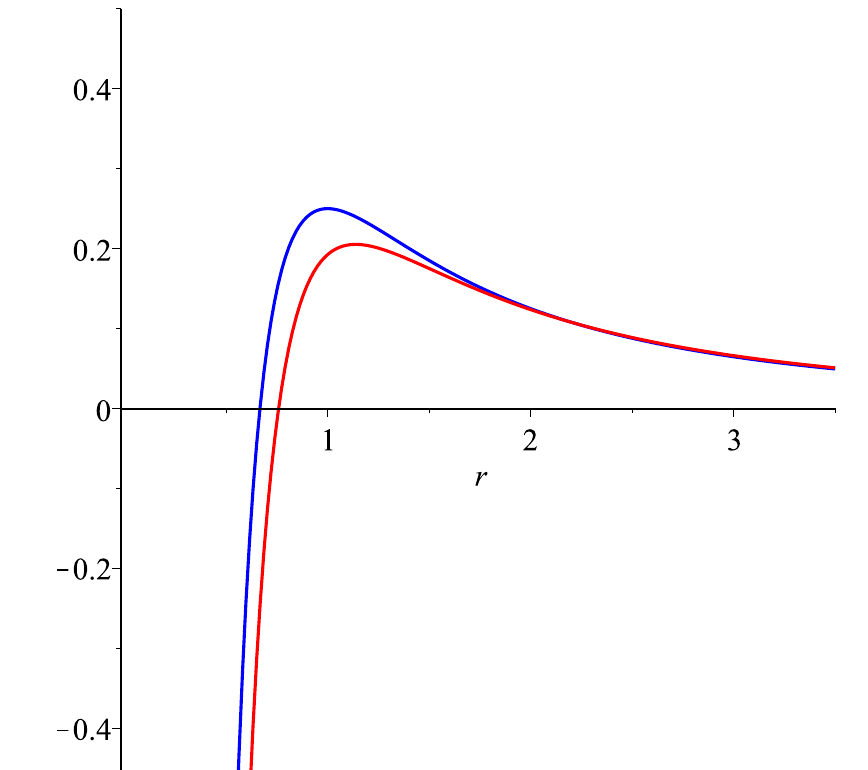} 
\caption{The plots of the equatorial effective potential of a non-extremal asymptotically flat Kerr black hole with generic parameters (bottom curve), and that of the extremal one (top curve). The extremal black hole has a photon orbit on its event horizon, but since it corresponds to a local maximum, it is unstable. \label{kerrplot}} 
\end{center}
\end{figure}

Consider the circular photon orbits, whose expressions are given in Eq.(\ref{1}). Evidently, in the extremal case, $a \to M$, we see that there are two photon orbits: a counter-rotating one at $r=4M$ and a co-rotating one on the horizon $r=M$. As we have mentioned in Sec.(1), there had been some debate regarding whether the co-rotating photon orbit really coincides with the event horizon. In any case, this orbit is an unstable one \cite{hobson}, which is clearly seen in the plot of the effective potential, as in Fig.(\ref{kerrplot}). Therefore, an extremal Kerr black hole does not admit a stable photon orbit. 

\subsection{Where Is the Photon Orbit of an Extremal Asymptotically Flat Kerr-Newman Black Hole?}

The Kerr-Newman black hole, which is both charged and rotating, is described by the metric tensor in Eq.(\ref{kerrmetric}), but with the mass term $2M/r$ replaced by $2M/r -Q^2/r^2$.
(Equatorial circular geodesics on asymptotically flat Kerr-Newman background were first studied in \cite{DK}. See also \cite{S1981, BBS}.) 

A straightforward generalization of the derivation in \cite{hobson} leads to the ``effective potential'' of massless particles for an asymptotically flat Kerr-Newman black hole on the equatorial plane\footnote{This has, incidentally, the same form of the effective potential for Kerr black hole in Eq.(\ref{K_V}), but with $2M/r \mapsto 2M/r -Q^2/r^2$. The same substitution also relates the effective potential of Schwarzschild black hole to that of Reissner-Nordstr\"om black hole in Eq.(\ref{RN_V}).} (See also Eq.(19) of \cite{1406.1295}.):
\begin{equation}
V(r,a,b)=\frac{J^2}{r^2}\left[1-\left(\frac{a}{b}\right)^2
-\left(\frac{2M}{r} - \frac{Q^2}{r^2} \right)\left(1-\frac{a}{b}\right)^2\right].
\label{KN_V}
\end{equation}

\begin{figure}
\centering
\includegraphics[width=3.472in]{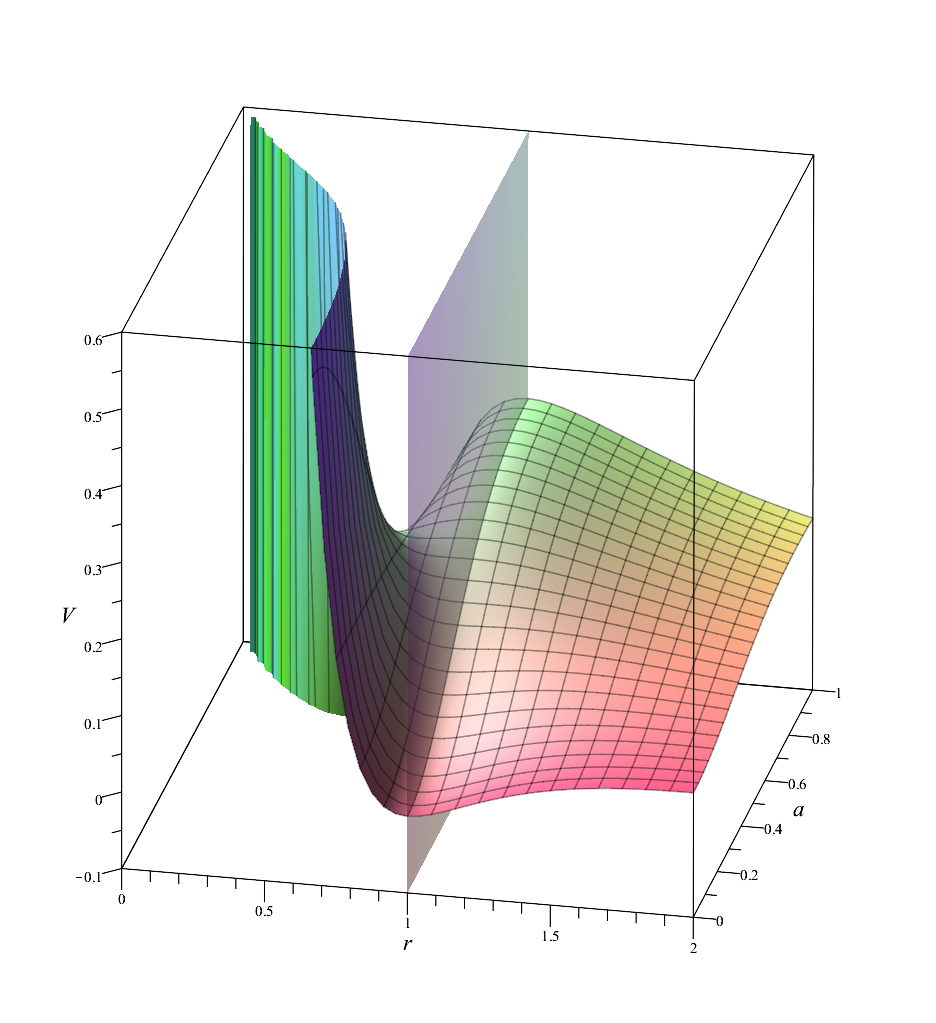}
\caption{The effective potential of an extremal asymptotically flat  Kerr-Newman black hole with unit mass. We also set the particle angular momentum $J=1$. The vertical plane at $r=1$ denotes the extremal horizon. Note that when $a=0$ the effective potential reduces to that of an extremal asymptotically flat  Reissner-Nordstr\"om spacetime, with a local minimum on the horizon. This gradually transits into a local maximum, and as $a=1$, recovers the result of an extremal asymptotically flat  Kerr spacetime. The turnaround point from local minimum to local maximum is $a=1/2$. In general, it is $a=M/2$.\label{kerrnewman}}
\end{figure}

Using the extremality conditions $M^2 = a^2 + Q^2$, one can find the value of the impact parameter necessary for equatorial geodesics of massless particles to be on the extremal horizon $r_h=M$. 
It is
\begin{equation}\label{b1}
b=\frac{2M^2-Q^2}{\sqrt{M^2-Q^2}}.
\end{equation}
For a consistency check, we note that setting $Q=0$ recovers the expression for an asymptotically flat Kerr black hole, namely that $b=2M$; whereas by taking $M \to Q$ limit we recover the extremal Reissner-Nordstr\"om case, in which the effective potential has no $b$-dependence. In fact, as shown in Fig.(\ref{kerrnewman}), the shape of the effective potential reduces correctly in both limits ($a=0$ and $a=1$).

We are now ready to study the photon orbit of an extremal Kerr-Newman black hole. Recall that the mystery is this:  in \cite{1503.01973}, Ulbricht and Meinel found that for an extremal asymptotically flat  Kerr-Newman black hole, the photon orbit only lies on the horizon if the angular momentum is sufficiently large $(a > M/2)$. How then would there be a photon orbit on the extremal RN horizon, which has $a = 0$? 

\begin{figure}
\centering
\includegraphics[width=3.2in]{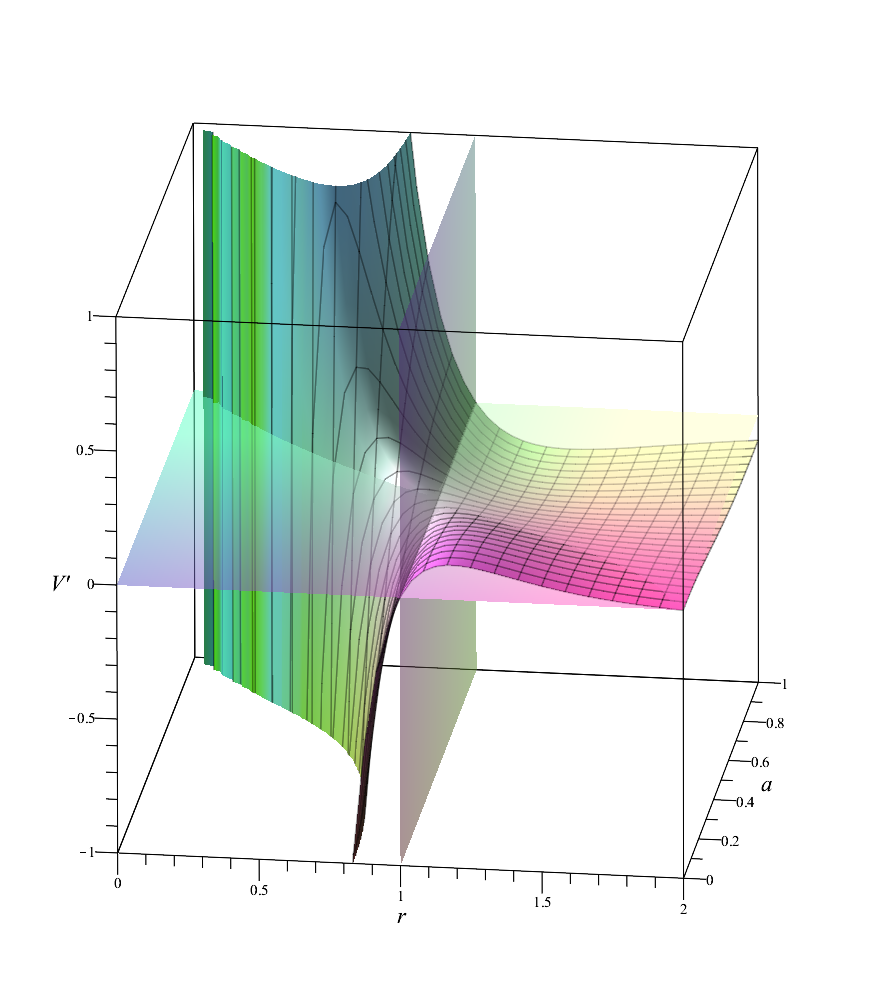}
\caption{The surface plot of the first derivative of the effective potential, $V'$, of an extremal asymptotically flat Kerr-Newman black hole, with mass $M$ set to unity, as a function of the angular momentum parameter $a$, and Boyer-Lindquist radius $r$. We also set the particle angular momentum $J=1$. The impact parameter $b$ has been set in accordance to Eq.(\ref{b1}). The charge is determined by the extremality condition $Q^2=M^2-a^2$.  Only the intersection of the surface with the vertical plane (the horizon) satisfies $V=1/b^2$, and is therefore the circular photon orbit with impact parameter given by Eq.(\ref{b1}).}\label{KNder}
\end{figure}

The answer is found by plotting the first derivative of the potential, $V'$, with $b$ set to the value in Eq.(\ref{b1}). 
Explicitly, we then have
\begin{equation}
V'(r,a)=\frac{-2J^2M^2[2a^2(r^2-M^2)+M^2(r^2-3Mr+2M^2)]}{r^5(M^2+a^2)^2}.
\end{equation}
The zeroes of $V'$ correspond to the possible positions of the photon orbits. The plot is given in Fig.(\ref{KNder}). 
The intersections of the surface with the horizontal $ar$-plane are the zeroes of $V'$, but they do not necessarily correspond to circular photon orbits, because one has to check that the condition $V(b,r)=1/b^2$ also holds.
The vertical plane corresponds to the position of the extremal event horizon. 
Note that the shape of $V' $ is such that \emph{there are always zeroes on the horizon}, these are genuine photon orbits. Ignoring what happens inside the horizon for the moment\footnote{At least formally, stable photon orbits are present inside the horizon for $a<M/2$. This is consistent with earlier works that show the existence of such peculiar orbits even for non-extremal cases \cite{calvani2}.}, for $a > M/2 $, these photon orbits on the horizon are unstable. Furthermore there is no other (co-rotating) photon orbit outside the horizon.

\begin{figure}
\centering
\includegraphics[width=3.5in]{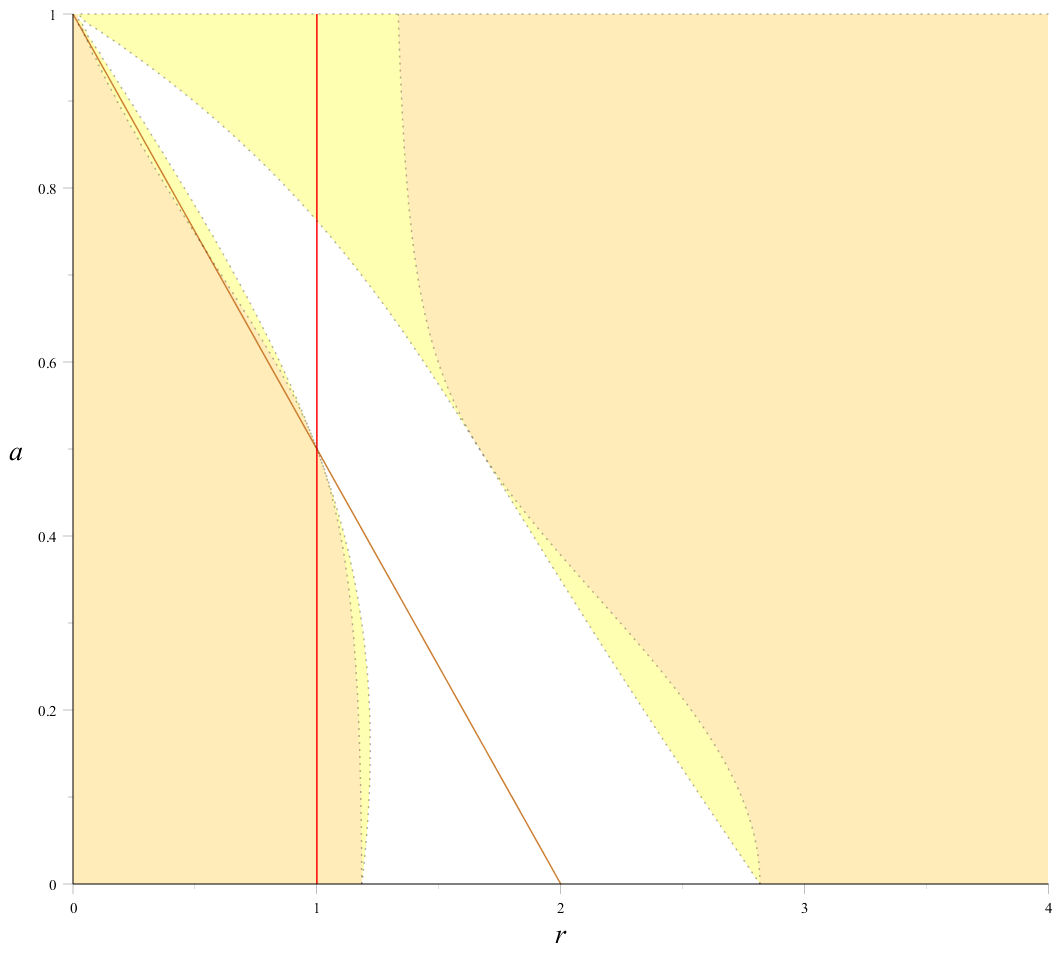}
\caption{This plot shows the values of $a$ and $r$ such that the first derivative of the effective potential, $V'$, vanishes, as well as satisfying $V=1/b^2$. That is, the solid slanted line and the solid vertical line (which coincides with the extremal horizon $r_h=1$) denote the circular photon orbits. The common shaded regions are where the function $\mathcal{G}(a,r)=(6a^2+3)r^2-12r+10(1-a^2)$ is positive. This function has the same sign as $V''$, and therefore governs the stability of the photon orbit \emph{on the horizon}. (The other photon orbit is shown for comparison purpose, its stability regions are slightly bigger, which include the lighter (yellow) parts. Retrograde photon orbits are not shown.) To conclude, the photon orbit on the horizon is stable if it is inside the common shaded region, and unstable otherwise.\label{secondder}}
\end{figure}

For $a < M/2$, the orbit on the horizon is stable (this can be shown analytically by examining $V''$, see below). But there is also \emph{another} photon orbit exterior to the horizon, which is unstable. This unstable photon orbit does not correspond to the other intersection of the surface with the horizontal $ar$-plane shown in Fig.(\ref{KNder}), as this curve does not satisfy the condition $V(b,r)=1/b^2$. Rather, by solving the equations $V(b,r)=1/b^2$ and $V'(b,r)=0$ simultaneously, we find that the other orbit satisfies $r=2(M-a)$, which corresponds to an impact parameter $b=-3a+4$.  In the $a \to 0$ limit, this outer photon orbit goes to $r=2M$. This recovers the result for asymptotically flat Reissner-Nordstr\"om black holes. We plotted both photon orbits in Fig.(\ref{secondder}).

In \cite{1503.01973}, the authors only tracked the outermost (co-rotating) photon orbit\footnote{Indeed, the authors of \cite{1503.01973} already derived the full result: immediately under Eq.(15) in their paper, they wrote ``We see that $\xi= a^2$ is a solution for prograde orbits. That is equivalent to the horizon's coordinate $x = 1$ of the extreme Kerr-Newman spacetime.'' This agrees with our result that there is always a photon orbit on the extremal horizon. However, in the subsequent discussion, they only discussed the outermost photon orbits (the outer solution to their polynomial Eq.(14)).}, and thus their result gives the impression that the unstable photon orbit on the extremal Kerr-Newman horizon ``peels away'' as rotation slows down to $a < M/2$. However, what actually happens is that it ``splits off'' into two  --- one moves outward and remains unstable as the black hole slows down to $a < M/2$, and one remains on the extremal horizon, but has become stable. Alternatively, one could say that the unstable orbit does indeed peel away from the extremal horizon, but a new stable orbit appears on the extremal horizon for $a < M/2$. Our result therefore agrees with that of \cite{1503.01973}, if we interpret their photon orbit to be the unstable one.

Let us show explicitly the stability analysis. For $M=1$ and $Q^2=1-a^2$, elementary calculus gives:
\begin{equation}
V''(r,a)=\frac{2J^2[(6a^2+3)r^2-12r+10(1-a^2)]}{r^6(1+a^2)^2}.
\end{equation}
The sign of $V''(a,r)$ therefore depends only on the quadratic polynomial $\mathcal{G}(a,r):=(6a^2+3)r^2-12r+10(1-a^2)$.
On the horizon $r_h=1$, this reduces to $\mathcal{G}(a,r_h)=1-4a^2$. Stability requires that $1-4a^2 > 0$, i.e., $a<1/2$. In general, it is $a<M/2$. The stability region of the photon orbit can be seen from Fig.(\ref{secondder}), where we have shaded the region in which the polynomial $\mathcal{G}(a,r)$ is positive. 

For the non-extremal case, there can be no photon orbit on the horizon. To see this, recall that the conditions for a circular photon orbit $r_h$ are\footnote{Note that the first condition is important when rotation is involved. In the static case, the effective potential has no $b$-dependence, so the value of $b$, and hence also of $E_\infty$, are trivially fixed by the solution of $V'(r)=0$.}
: $V(r_h,a,b)=E_\infty^2$ and that $V'(r_h,a,b)=0$. 
(One can check that with the choice of $b$ in Eq.(\ref{b1}), the extremal black hole discussed above does indeed satisfy these two conditions.)
The first condition is, explicitly
\begin{equation}
\frac{1}{r^2}\left(1-\frac{a}{b}\right)\left[1+\frac{a}{b} -\left(\frac{2M}{r}-\frac{Q^2}{r^2}\right)\left(1-\frac{a}{b}\right)\right] = \frac{1}{b^2}.
\end{equation}
Note that $E_\infty^2=J^2/b^2$, and $J^2$ cancels out since $V$ also has a factor of $J^2$.
This leads to the algebraic equation
\begin{equation}\label{cond1}
[r^2(b+a)-2Mr(b-a)+Q^2(b-a)](b-a)=r^4.
\end{equation}
The second condition $V'(r_h,a,b)=0$ yields
\begin{equation}
[-r^2(b+a)+3Mr(b-a)-2Q^2(b-a)](b-a)=0.
\end{equation}
Since $a\neq b$ --- for otherwise the potential $V$ vanishes identically --- we must have 
\begin{equation}\label{cond2}
b-a = \frac{r^2(b+a)}{3Mr-2Q^2}.
\end{equation}
Substituting Eq.(\ref{cond2}) into Eq.(\ref{cond1}), elementary algebra yields
\begin{equation}
b = \frac{r^2}{\sqrt{Mr-Q^2}} + a.
\end{equation}
Substituting this expression back into Eq.(\ref{cond2}) then gives
\begin{equation}
b=\frac{3Mr-2Q^2}{\sqrt{Mr-Q^2}}-a.
\end{equation}
Equating these two expressions for $b$, we get
\begin{equation}\label{cond3}
3Mr-2Q^2-r^2=2a\sqrt{Mr-Q^2}.
\end{equation}
Now suppose that the extremal event horizon $r_h=M+\sqrt{M^2-Q^2-a^2}$ is a solution, i.e., that it is a photon orbit. Then 
\begin{equation}
r_h^2 = 2Mr-Q^2 - a^2,
\end{equation}
and consequently Eq.(\ref{cond3}) reduces to a perfect square
\begin{equation}
(a-\sqrt{Mr_h-Q^2})^2 = 0.
\end{equation}
Subsequently,
\begin{equation}
a^2 = Mr_h- Q^2 \Rightarrow (M^2-Q^2-a^2)^2 = M^2(M^2 - Q^2 -a^2),
\end{equation}
which implies
\begin{equation}
(Q^2+a^2)(M^2-Q^2-a^2)=0.
\end{equation}
We must therefore have $M^2-Q^2-a^2=0$, i.e., the black hole is necessarily extremal. Note that $Q^2+a^2$ cannot be zero unless $Q=0=a$, which means the black hole is Schwarzschild. However, recall that the photon orbit for a Schwarzschild black hole is located at $r=3M$, \emph{not} on its horizon.

\section{Conclusion: Light on the Event Horizon of Extremal Black Holes}

Extremal black holes have always been a curiosity in general relativity. Their virtues lie in their simplicities. Due to the extremality condition that relates the black hole parameters, many equations become much easier to solve. Extremal black holes are also of great interest in theoretical physics, since they do not emit Hawking radiation (though they can discharge via Schwinger emission. See, for example, \cite{khrip}). Due to the subtleties involved in taking limits of spacetimes, the properties of an exactly extremal black hole can differ from a near-extremal one, however close to extremality the latter is. In particular, it has been shown in the literature that many extremal black holes admit photon orbits \emph{exactly} on their event horizons. Sometimes such an orbit is not stable, but surprisingly, \emph{sometimes it is}. It is therefore an interesting question to ask: when is a photon orbit on the extremal horizon stable?

In this work we have shown that, if the extremal black hole is static and spherically symmetric, then the photon orbit on its horizon is necessarily stable if the first nonzero Taylor coefficient of its metric coefficient $f^{(k)}(r_h)/k!$, $k \geqslant 2$, is with an even $k$. This is a rather surprising property that was first discovered in the context of asymptotically flat Reissner-Nordstr\"om black holes \cite{1001.0359}. Though the result does not extend straightforwardly to rotating black holes in general, it is still rather amazing that light can orbit so close to a black hole --- on the horizon itself --- without falling into the hole. 

The rotating case is much more complicated to analyze, but is nevertheless important to understand since astrophysical black holes do rotate. In addition, rotating (and shearing) black holes play important roles in holography when the dual field theory exhibits nonzero angular momentum, e.g., in the context of heavy ion collisions \cite{1206.0120, 1403.3258}. It should be emphasized that in the Boyer-Lindquist coordinates, although the extremal Kerr and Kerr-Newman black holes have a photon orbit at $r=r_h$, the photon orbit does not really coincide with the event horizon, but is located at a finite proper distance away. This is due to the degeneracy of the radial coordinate as the Kerr(-Newman) throat geometry tends to that of a cylinder in the extremal limit. However, the event horizon does coincide with the photon orbit in the Doran frame \cite{1107.5081}. 

We examined the extremal asymptotically flat  Kerr and Kerr-Newman black holes, and found that they always have photon orbit on the equatorial plane of their event horizon. For the Kerr case this orbit is unstable, whereas for the Kerr-Newman case, it is stable if $a < M/2$ --- or equivalently $Q^2 > 3M^2/4$ --- but unstable if $a >M/2$. That is to say, angular momentum tends to destabilize the photon orbit, whereas electrical charge tends to stabilize it. We also found that, for $a < M/2$, in addition to the photon orbit on the event horizon, there appears another unstable photon orbit which lies outside of the horizon. This clarifies the --- at least at first sight --- seemingly contradictory results in the literature. We also prove that there cannot be any photon orbit on the event horizon of a non-extremal Kerr-Newman black hole. Perhaps some further understanding can be gained by investigating the stability of photon orbits on the horizon of an extremal asymptotically flat  Kerr-Newman black hole but without restricting the analysis to the equatorial plane. Specifically, it would be interesting to see how stability condition might change with latitude, away from the equatorial plane, as the value of the rotation parameter is increased.

In the theorem just before Sec.(\ref{sss}),  we characterized the stability of photon orbits for spherically symmetric black hole spacetimes. Essentially, if the metric coefficient is analytic, then this depends on the sign of the first nonzero Taylor coefficient of the expansion. As we have mentioned, though unlikely, it is conceivable that a photon orbit might have vanishing second derivative $f''(r_h)=0$, i.e., the photon orbit corresponds to an inflection point of the potential. Then the orbit is ``one-sided unstable''. We know of a concrete example of a photon orbit that lies on the inflection point of the potential: it is not a black hole but an asymptotically flat Reissner-Nordstr\"om naked singularity. One can easily show, from the potential in Eq.(\ref{RN_V}), that for a naked singularity $(Q > M)$, we still have a stable photon orbit as long as the charge-to-mass ratio is not too high. The limiting case is when $Q^2/M^2 = 9/8$, with the photon orbit at $r_{\text{ph}}=3M/2$. The photon orbit changes from being a local minimum (for $Q^2/M^2 < 9/8$) to an inflection point (at $Q^2/M^2 =9/8$), and stability is lost. The point is that the stability of photon orbit is definitely not a feature unique to an extremal horizon, and if naked singularity exists, then stable photon orbits can exist at least in some range of the parameters of the theory.

An open problem is the following: In this paper we only studied fairly simple spacetimes which are either spherically symmetric or axially symmetric, what happens when we have an extremal black hole with less symmetries? Does an extremal black hole always have a photon orbit on its horizon? More importantly, under what conditions would horizon photon orbits be stable? Of course, given the metric of any extremal black hole, we could always analyze the photon trajectories on a case by case basis, but it would be better if one could make a stronger statement. Unfortunately at this stage our analysis does not permit us to say anything definitive about the more general spacetimes that have less symmetry. It is, however, quite clear that \emph{some} kind of symmetry would be required to have a photon orbit. Even the existence of spherical orbits (i.e. at constant $r$) --- not necessarily the one on $r_h$ --- in asymptotically flat Kerr background, is a subtle issue once one studies motions that are spherical but not restricted to the equatorial plane \cite{teo}. In that case massless particle would have some periodic motion in the latitudinal direction, and this is governed by a hidden symmetry, quantified by Carter's constant \cite{HCR1}. This symmetry itself is not obvious from the spacetime symmetry alone. This suggests that one should first study if a given geometry has enough symmetry that governs the geodesics, so that at least in principle, a photon orbit could exist. This task is, unfortunately, also not a trivial one, especially when the symmetry is hidden.

Next, we remark that the \emph{instability} of a photon orbit is crucial to guarantee the \emph{stability} of a given spacetime. The reason is the following: if the photon orbit is stable, one could pile up arbitrarily many photons on said orbit. (This would also be the case for gravitons and other massless particles.) This would eventually lead to a backreaction to the spacetime geometry. That is, the geometry will deform away from the one described by the original metric. In fact, the instability of the photon orbit for Kerr black holes outside their horizon was a crucial ingredient used in the proof that slowly rotating Kerr black holes are stable when considering linear wave equation on the background\cite{AB}. The fact that static extremal black holes have stable photon orbits, on the contrary, implies that these black holes are unstable (even without considering the Aretakis instability\cite{1110.2006, 1110.2007, 1110.2009, 1206.6598, 1208.1437, 1211.6903, 1307.6800}). The question is then: what exactly happens to the geometry due to this backreaction? It is natural to postulate that they would settle down to a ``nearby solution'', perhaps a non-extremal black hole. However, one would need to actually show this.

Finally, we remark that in the Kerr/CFT correspondence \cite{livrev,1103.2355}, it is noted that since the event horizon of the extremal Kerr black hole rotates at the speed of light, heuristically this means that physical objects must also move at the speed of light if it were to stay on the horizon (so only massless fields are allowed), and hence only chiral degrees of freedom appear \cite{livrev,1103.2355}, which are responsible for the entropy of the black hole. So an extremal Kerr black hole is dual to the chiral limit of a two-dimensional CFT.
We wonder whether the stability --- or in the extremal Kerr case, instability --- of the photon orbit on the extremal horizon has any implication to the field theory correspondence.  In addition, although we have investigated the photon orbits in quite general spacetimes for the static case, we have only investigated asymptotically flat Kerr and Kerr-Newman black holes in the non-static case. It would therefore be interesting to see if in the asymptotically AdS case, the photon orbits of Kerr-Newman black holes behave in a similar way (it is already known that a stable photon orbit exists inside the inner horizon of an asymptotically anti-de Sitter Kerr-Newman black hole \cite{4500311}) and what --- if any --- implication this has on holography \cite{1009.1661}.

\section*{Acknowledgements}

F.S. Khoo gratefully acknowledges the grant support from DFG RTG-Models of Gravity. Y.C. Ong thanks Brett McInnes and Keisuke Izumi for useful discussions, comments, and suggestions. Y.C. Ong also thanks the Riemann Center for Geometry and Physics, at the Leibniz Universit\"at Hannover, for the Riemann fellowship and hospitality during which this paper was finalized, as well as Nordita for travel support.

\section*{Appendix}
\noindent
The metric tensor of an $(n+2)$-dimensional static topological black hole is
\begin{equation}
g [\text{AdSRN}_{n+2}^k]
= - f(r) dt^2 + f(r)^{-1} dr^2 + r^2 d\Omega^2[X_n^k] \, ,
\end{equation}
for $n \geqslant 2$,
where
\begin{equation}
f(r) = k + \frac{r^2}{L^2} - \frac{16 \pi M}{n \Gamma[X^k_n] r^{n-1}} 
+ \frac{8\pi Q^2}{n(n-1) (\Gamma[X^k_n])^2 r^{2n-2}} \, . 
\end{equation}
Here $\Gamma [X^k_n]$ is a constant dimensionless area of an $n$-dimensional Riemannian manifold $X_n^k$ with curvature $k = -1, 0, 1$, while
$L$ is the AdS curvature length scale. \newline
\\
Let
\begin{equation}
f(r) = k + \frac{r^2}{L^2} - \frac{A}{r^{n-1}} + \frac{B}{r^{2(n-1)}} \, ,
\end{equation}
where 
\begin{equation}
A := \frac{16 \pi M}{n \Gamma[X^k_n]} \hspace{2cm}
\text{and} \hspace{2cm}
B := \frac{8 \pi Q^2}{n(n-1) (\Gamma[X^k_n])^2} \,.
\end{equation}
We find that
\begin{equation}
f'(r) = \frac{2r}{L^2} + \frac{(n-1)A}{r^n} - \frac{2(n-1)B}{r^{2n-1}} \, .
\end{equation}
Extremality means that
\begin{equation}
f(r_h) = 0 = f'(r_h) \,.
\end{equation}
From $f(r_h) = 0$, we have
\begin{equation}
k r_h^{2n-1} + \frac{r_h^{2n+1}}{L^2} - A r_h^n + Br_h = 0,
\label{fr}
\end{equation}
and from $f'(r_h) = 0$, we have
\begin{equation}
\frac{2 r_h^{2n+1}}{L^2} + (n-1) A r_h^n - 2(n-1)Br_h = 0 \, .
\label{f'r}
\end{equation}
Substituting (\ref{fr}) into (\ref{f'r}), we get 
\begin{equation}
(n+1) A r_h^n = 2n B r_h + 2kr_h^{2n-1} \, .
\label{c1}
\end{equation} 
We can read off the constants
\begin{equation}
A = \frac{2r_h(nB + k r_h^{2n-2})}{(n+1)r_h^n} \, ,
\label{A}
\end{equation}
\begin{equation}
B = \frac{(n+1) A r_h^n}{2n r_h} - \frac{k r_h^{2n-2}}{n} \, .
\label{B}
\end{equation}
Substituting (\ref{A}) back into (\ref{fr}), we obtain
\begin{eqnarray}
\frac{r_h^{2n+1}}{L^2} 
&=& \frac{2r_h(nB + k r_h^{2n-2})}{n+1} - Br_h - kr_h^{2n-1}
\\
&=& \frac{n-1}{n+1} (Br_h - k r_h^{2n-1} )\, . 
\end{eqnarray}
Hence,
\begin{equation}
\frac{r_h^{2n}}{L^2} = \frac{n-1}{n+1} ( B - kr_h^{2n-2} )\, .
\label{c2}
\end{equation} 
From the substitution of (\ref{B}) into (\ref{c2}), we obtain 
\begin{equation}
\frac{r_h^{2n}}{L^2} = \frac{n-1}{2n} A r_h^{n-1} + \frac{(1-n)}{n} k r_h^{2n-2},
\end{equation}
and thus
\begin{equation}
\frac{r_h^{n+1}}{L^2} = \frac{n-1}{2n} A + \frac{(1-n)}{n} k r_h^{n-1} \, .
\label{arh}
\end{equation} 
%The equation describing a photon with energy $E_{\infty}$ along its orbit is
%\begin{equation}
%\dot{r}^2 + V(r) = E_{\infty}^2 \, ,
%\end{equation}
%where
%\begin{equation}
%V(r) = \frac{J^2 f(r)}{r^2} \, .
%\end{equation}
For the effective potential which corresponds to a photon orbit in the spacetime that we are interested in,
\begin{equation}
V(r) = \frac{J^2}{r^2} 
\left( k + \frac{r^2}{L^2} - \frac{A}{r^{n-1}} + \frac{B}{r^{2(n-1)}} \right) \, .
\end{equation}
So we find
\begin{equation}
\frac{V'(r)}{J^2} = -\frac{2k}{r^3} + 
\frac{(n+1)A}{r^{n+2}} - \frac{2nB}{r^{2n+1}} \, ,
\end{equation}
and consequently,
\begin{equation}
\frac{V''(r)}{J^2} =
\frac{6k}{r^4} -
\frac{(n+1)(n+2)A}{r^{n+3}} + \frac{2n(2n+1)B}{r^{2(n+1)}} \, .
\label{finalV''}
\end{equation}
Inserting (\ref{B}) in (\ref{finalV''}) at the extremal horizon $r = r_{\text{h}}$ gives
\begin{eqnarray}
V''(r_h) 
&=& 
J^2 \left[ \frac{6k}{r_h^4} + \frac{A}{r^{n+3}_h} \, (n^2-1) - \frac{2k(2n+1)}{r^4_h} \right]
\\
&=& J^2 \left[ \frac{A}{r^{n+3}_h} \, (n^2-1) + \frac{4k}{r_h^4} (1-n) \right] \, .
\label{11}
\end{eqnarray} 
By using (\ref{arh}), we simplify (\ref{11}) to
\begin{eqnarray}
V''(r_h) 
&=& J^2 \left[\frac{A (n^2 - 1)}{r_h^{n+3}} + \frac{4n}{r_h^2 L^2} - \frac{2A(n-1)}{r_h^{n+3}}\right]
\\
&=& 
J^2\left[\frac{4n}{r_h^2 L^2} + \frac{A (n -1)^2}{r_h^{n+3}}\right] \, .
\end{eqnarray}
Since the dimensionality $n$ is at least 2, $V''(r_h)$ is positive. 
Therefore, the extremal photon orbit is stable.

%%%%%%%%%%%%%%%%%%%%%%%%%%%%%%%%%%%%%%%%%%%%%%%%%%%%%%%%%%%%%%%%%%%%%%%%%%%%%%%%%%%%%%%%%%%%%%%%%%%%%%%%%%%%%%%%%%%%%%%%%%%%%%%%%%%%%%%%%%%%%%%%%%%%%%%%%%


\begin{thebibliography}{99}

\bibitem{wald}
Robert Wald, \emph{General Relativity}, The University of Chicago Press, Chicago and London, 1984.

\bibitem{raine}
Derek Raine, Edwin Thomas, \emph{Black Holes: A Student Text}, Third Edition, Imperial College Press, 2015.

\bibitem{BPT}
James M. Bardeen, William H. Press, Saul A. Teukolsky, ``Rotating Black Holes: Locally
Nonrotating Frames, Energy Extraction, and Scalar Synchrotron Radiation'', {\changeurlcolor{vividviolet}\href{http://adsabs.harvard.edu/full/1972ApJ...178..347B}{Astrophys. J. \textbf{178} (1972) 347}.} 



\bibitem{1210.2486}
Shahar Hod, ``Spherical Null Geodesics of Rotating Kerr Black Holes'', {\changeurlcolor{vividviolet}\href{http://www.sciencedirect.com/science/article/pii/S0370269312013007}{Phys. Lett. B \textbf{718} (2013) 1552}}, \href{http://arxiv.org/abs/1210.2486}{[arXiv:1210.2486 [gr-qc]]}.

\bibitem{teo}
Edward Teo, ``Spherical Photon Orbits Around a Kerr Black Hole'', {\changeurlcolor{vividviolet}\href{http://link.springer.com/article/10.1023\%2FA\%3A1026286607562}{Gen. Rel. Grav. \textbf{35} (2003) 1909}}.


\bibitem{AL}
Marek A. Abramowicz, Jean-Pierre Lasota, ``A Note on a Paradoxical Property of the Schwarzschild Solution'', {\changeurlcolor{vividviolet}\href{http://www.actaphys.uj.edu.pl/fulltext?series=Reg&vol=5&page=327}{Acta Phys. Polonica. \textbf{B5} (1974) 327}}.


\bibitem{AP}
Marek A. Abramowicz, Aragam Prasanna, ``Centrifugal-Force Reversal Near a Schwarzschild Black Hole'', {\changeurlcolor{vividviolet}\href{http://adsabs.harvard.edu/full/1990MNRAS.245..720A}{MNRAS \textbf{245} (1990) 720}}. 

\bibitem{Rickard}
Rickard Jonsson, ``An Intuitive Approach to Inertial Forces and the Centrifugal Force Paradox in General Relativity'', {\changeurlcolor{vividviolet}\href{http://scitation.aip.org/content/aapt/journal/ajp/74/10/10.1119/1.2198880}{Am. J. Phys. \textbf{74} (2006) 905}}, \href{http://arxiv.org/abs/0708.2488}{[arXiv:0708.2488 [gr-qc]]}. 

\bibitem{BBI}
Claude Barrab\`es, Bruno Boisseau, Werner Israel, ``Orbits, Forces and Accretion Dynamics near Spinning Black Holes'', {\changeurlcolor{vividviolet}\href{http://mnras.oxfordjournals.org/content/276/2/432}{Mon. Not. Roy. Astron. Soc. \textbf{276} (1995) 432}}, \href{http://arxiv.org/abs/gr-qc/9505025}{[arXiv:gr-qc/9505025]}.

\bibitem{1406.5475}
Carla Cederbaum, ``Uniqueness of Photon Spheres in Static Vacuum Asymptotically Flat Spacetimes'', \href{http://arxiv.org/abs/1406.5475}{[arXiv:1406.5475 [math.DG]]}.

\bibitem{1504.05804}
Carla Cederbaum, Gregory J. Galloway, ``Uniqueness of Photon Spheres via Positive Mass Rigidity'', \href{http://arxiv.org/abs/1504.05804}{[arXiv:1504.05804 [math.DG]]}.


\bibitem{takahashi}
Rohta Takahashi, ``Shapes and Positions of Black Hole Shadows in Accretion Disks and Spin Parameters of Black Holes'',  {\changeurlcolor{vividviolet}\href{https://iopscience.iop.org/article/10.1086/422403/meta}{Astrophys. J. \textbf{611} (2004) 996}}, \href{http://arxiv.org/abs/astro-ph/0405099v1}{[arXiv:astro-ph/0405099]}.


\bibitem{EHT1}
Ru-Sen Lu, Avery E. Broderick, Fabien Baron, John D. Monnier, Vincent L. Fish, Sheperd S. Doeleman, Victor Pankratius, ``Imaging the Supermassive Black Hole Shadow and Jet Base of M87 with the Event Horizon Telescope'', {\changeurlcolor{vividviolet}\href{https://iopscience.iop.org/article/10.1088/0004-637X/788/2/120/meta}{Astrophys. J. \textbf{788} (2014) 120}}, \href{http://arxiv.org/abs/1404.7095}{[arXiv:1404.7095 [astro-ph.IM]]}. 

\bibitem{EHT2}
Dimitrios Psaltis, Feryal Ozel, Chi-Kwan Chan, Daniel P. Marrone, ``A General Relativistic Null Hypothesis Test with Event Horizon Telescope Observations of the Black-Hole Shadow in Sgr A*'',  {\changeurlcolor{vividviolet}\href{https://iopscience.iop.org/article/10.1088/0004-637X/814/2/115/meta}{Astrophys. J. \textbf{814 }(2015) 2, 115}}, \href{http://arxiv.org/abs/1411.1454}{[arXiv:1411.1454 [astro-ph.HE]]}.

\bibitem{1503.01840}
Hiromi Saida, Atsuhito Fujisawa, Chul-Moon Yoo, Yasusada Nambu, ``Spherical Polytropic Balls Cannot Mimic Black Holes'', {\changeurlcolor{vividviolet}\href{http://ptep.oxfordjournals.org/content/2016/4/043E02}{PTEP \textbf{2016} (2016) 4, 043E02 }}, \href{http://arxiv.org/abs/1503.01840}{[arXiv:1503.01840 [gr-qc]]}.

\bibitem{1410.1894}
M. C. Baldiotti, Walace S. Elias, Carlos Molina, Thiago S. Pereira, ``Thermodynamics of Quantum Photon Spheres'', {\changeurlcolor{vividviolet}\href{http://journals.aps.org/prd/abstract/10.1103/PhysRevD.90.104025}{Phys. Rev. D \textbf{90} (2014) 104025}}, \href{http://arxiv.org/abs/1410.1894}{[arXiv:1410.1894 [gr-qc]]}.

\bibitem{page}
Don N. Page, ``Particle Emission Rates from a Black Hole: Massless Particles from an Uncharged, Nonrotating Hole'', {\changeurlcolor{vividviolet}\href{http://journals.aps.org/prd/abstract/10.1103/PhysRevD.13.198}{Phys. Rev. D \textbf{13} (1976) 198}}.

\bibitem{1602.03837}
The LIGO Scientific Collaboration, the Virgo Collaboration, ``Observation of Gravitational Waves from a Binary Black Hole Merger'', {\changeurlcolor{vividviolet}\href{http://journals.aps.org/prl/abstract/10.1103/PhysRevLett.116.061102}{Phys. Rev. Lett. \textbf{116} (2016) 061102}}, \href{http://arxiv.org/abs/1602.03837}{[arXiv:1602.03837 [gr-qc]]}.

\bibitem{1602.07309}
Vitor Cardoso, Edgardo Franzin, Paolo Pani, ``Is the Gravitational-Wave Ringdown a Probe of the Event Horizon?'', {\changeurlcolor{vividviolet}\href{http://journals.aps.org/prl/abstract/10.1103/PhysRevLett.116.171101}{Phys. Rev. Lett. \textbf{116} (2016) 171101}}, \href{http://arxiv.org/abs/1602.07309v1}{[arXiv:1602.07309 [gr-qc]]}.

\bibitem{extreme}
Jeffrey E. McClintock, Ramesh Narayan, Shane W. Davis, Lijun Gou, Akshay Kulkarni, Jerome A. Orosz, Robert F. Penna, Ronald A. Remillard, James F. Steiner, ``Measuring the Spins of Accreting Black Holes'',  {\changeurlcolor{vividviolet}\href{https://iopscience.iop.org/article/10.1088/0264-9381/28/11/114009/meta}{Class. Quant. Grav. \textbf{28} (2011) 114009}}, \href{http://arxiv.org/abs/1101.0811}{[arXiv:1101.0811 [astro-ph.HE]]}.

\bibitem{extreme1}
Akshay K. Kulkarni, Robert F. Penna, Roman V. Shcherbakov, James F. Steiner, Ramesh Narayan, Aleksander Sadowski, Yucong Zhu, Jeffrey E. McClintock, Shane W. Davis, Jonathan C. McKinney, ``Measuring Black Hole Spin by the Continuum-Fitting Method: Effect of Deviations from the Novikov-Thorne Disc Model'', {\changeurlcolor{vividviolet}\href{http://mnras.oxfordjournals.org/content/414/2/1183}{MNRAS \textbf{414} (2) (2011) 1183}}, \href{https://arxiv.org/abs/1102.0010}{[arXiv:1102.0010 [astro-ph.HE]]}.


\bibitem{extreme2}
Lijun Gou, Jeffrey E. McClintock, Mark J. Reid, Jerome A. Orosz, James F. Steiner, Ramesh Narayan, Jingen Xiang, Ronald A. Remillard, Keith A. Arnaud, Shane W. Davis,
``The Extreme Spin of the Black Hole in Cygnus X-1'', {\changeurlcolor{vividviolet}\href{https://iopscience.iop.org/article/10.1088/0004-637X/742/2/85/meta;jsessionid=3905E7BB4079E04848564DBCF4FF9746.c3.iopscience.cld.iop.org#}{The Astrophysical Journal \textbf{742} (2011) 85}}, \href{https://arxiv.org/abs/1106.3690}{[arXiv:1106.3690 [astro-ph.HE]]}.

\bibitem{1110.2006}
Stefanos Aretakis, ``Decay of Axisymmetric Solutions of the Wave Equation on Extreme Kerr Backgrounds'', {\changeurlcolor{vividviolet}\href{http://www.sciencedirect.com/science/article/pii/S0022123612003114}{J. Functional Analysis \textbf{263} (2012) 2770}}, \href{http://arxiv.org/abs/1110.2006}{[arXiv:1110.2006 [gr-qc]]}.

\bibitem{1110.2007}
Stefanos Aretakis, ``Stability and Instability of Extreme Reissner-Nordstr\"{o}m Black Hole Spacetimes for Linear Scalar Perturbations I'', 	{\changeurlcolor{vividviolet}\href{http://link.springer.com/article/10.1007\%2Fs00220-011-1254-5}{Comm. Math. Phys. \textbf{307} (2011) 17}}, \href{http://arxiv.org/abs/1110.2007}{[arXiv:1110.2007 [gr-qc]]}.

\bibitem{1110.2009}
Stefanos Aretakis, ``Stability and Instability of Extreme Reissner-Nordstr\"{o}m Black Hole Spacetimes for Linear Scalar Perturbations II'', {\changeurlcolor{vividviolet}\href{http://link.springer.com/article/10.1007\%2Fs00023-011-0110-7}{Ann. Henri Poincare \textbf{8} (2011) 1491}}, \href{http://arxiv.org/abs/1110.2009}{[arXiv:1110.2009 [gr-qc]]}.

\bibitem{1206.6598}
Stefanos Aretakis, ``Horizon Instability of Extremal Black Holes'', \href{http://arxiv.org/abs/1206.6598}{[arXiv:1206.6598 [gr-qc]]}.

\bibitem{1208.1437}
James Lucietti, Harvey S. Reall, ``Gravitational Instability of an Extreme Kerr Black Hole'', {\changeurlcolor{vividviolet}\href{http://journals.aps.org/prd/abstract/10.1103/PhysRevD.86.104030}{Phys. Rev. D \textbf{86} (2012) 104030}}, \href{http://arxiv.org/abs/1208.1437}{[arXiv:1208.1437 [gr-qc]]}.

\bibitem{1211.6903}
Keiju Murata, ``Instability of Higher Dimensional Extreme Black Holes'', {\changeurlcolor{vividviolet}\href{https://iopscience.iop.org/article/10.1088/0264-9381/30/7/075002/meta}{Class. Quant. Grav. \textbf{30} (2013) 075002}}, \href{http://arxiv.org/abs/1211.6903}{[arXiv:1211.6903 [gr-qc]]}.

\bibitem{1307.6800}
Keiju Murata, Harvey S. Reall, Norihiro Tanahashi, ``What Happens at the Horizon(s) of an Extreme Black Hole?'', {\changeurlcolor{vividviolet}\href{https://iopscience.iop.org/article/10.1088/0264-9381/30/23/235007/meta}{Class. Quant. Grav. \textbf{30} (2013) 235007}}, \href{http://arxiv.org/abs/1307.6800}{[arXiv:1307.6800 [gr-qc]]}. 



\bibitem{carroll}
Sean M. Carroll, Matthew C. Johnson, Lisa Randall, ``Extremal Limits and Black Hole Entropy'', {\changeurlcolor{vividviolet}\href{https://iopscience.iop.org/article/10.1088/1126-6708/2009/11/109/meta}{JHEP \textbf{0911} (2009) 109}}, \href{http://arxiv.org/abs/0901.0931}{[arXiv:0901.0931 [hep-th]]}.

\bibitem{ingemar}
Ingemar Bengtsson, S\"oren Holst, Emma Jakobsson, ``Classics Illustrated: Limits of Spacetimes'', 	{\changeurlcolor{vividviolet}\href{https://iopscience.iop.org/article/10.1088/0264-9381/31/20/205008/meta}{Class. Quantum Grav. \textbf{31} (2014) 205008}}, \href{http://arxiv.org/abs/1406.4326}{[arXiv:1406.4326 [gr-qc]]}.

\bibitem{stotyn}
Sean Stotyn, ``A Tale of Two Horizons'', {\changeurlcolor{vividviolet}\href{http://www.nrcresearchpress.com/doi/10.1139/cjp-2015-0091}{Can. J. Phys. \textbf{93} (2015) 9, 995}}, \href{http://arxiv.org/abs/1502.02737}{[arXiv:1502.02737 [gr-qc]]}.

\bibitem{1001.0359}
Parthapratim Pradhan, Parthasarathi Majumdar, ``Circular Orbits in Extremal Reissner-Nordstr\"{o}m Spacetimes'', {\changeurlcolor{vividviolet}\href{http://www.sciencedirect.com/science/article/pii/S0375960110014635}{Phys. Lett. A \textbf{375} (2011) 474}}, \href{http://arxiv.org/abs/1001.0359v2}{[arXiv:1001.0359 [gr-qc]]}.

\bibitem{1108.2333}
Parthapratim Pradhan, Parthasarathi Majumdar, ``Extremal Limits and Kerr Spacetime'', {\changeurlcolor{vividviolet}\href{http://link.springer.com/article/10.1140\%2Fepjc\%2Fs10052-013-2470-2}{Eur. Phys. J. C \textbf{73} (2013) 2470}}, \href{http://arxiv.org/abs/1108.2333}{[arXiv:1108.2333 [gr-qc]]}.

\bibitem{podolsky}
Jiri Podolsk\'y, ``The Structure of the Extreme Schwarzschild-de Sitter Space-Time'', {\changeurlcolor{vividviolet}\href{http://link.springer.com/article/10.1023\%2FA\%3A1026762116655}{Gen. Rel. Grav. \textbf{31} (1999) 1703}}, \href{http://arxiv.org/abs/gr-qc/9910029v1}{[arXiv:gr-qc/9910029]}.

\bibitem{1210.0221}
Parthapratim Pradhan, ``Horizon Straddling ISCOs in Spherically Symmetric String Black Holes'',  {\changeurlcolor{vividviolet}\href{http://www.worldscientific.com/doi/10.1142/S0218271815500868}{Int. J. Mod. Phys. D \textbf{24} (2015) 1550086}}, \href{http://arxiv.org/abs/1210.0221}{[arXiv:1210.0221 [gr-qc]]}.

\bibitem{0803.2539}
Zden\v{e}k Stuchl\'ik, Stanislav Hled\'ik, ``Equatorial Photon Motion in the Kerr-Newman Spacetimes with a Non-Zero Cosmological Constant'', {\changeurlcolor{vividviolet}\href{https://iopscience.iop.org/article/10.1088/0264-9381/17/21/312}{Class. Quant. Grav. \textbf{17} (2000) 4541}}, \href{https://arxiv.org/abs/0803.2539}{[arXiv:0803.2539 [gr-qc]]}.

\bibitem{0803.2685}
Zden\v{e}k Stuchl\'ik, Stanislav Hled\'ik, ``Properties of the Reissner-Nordstr\"om Spacetimes with a Nonzero Cosmological Constant'', {\changeurlcolor{vividviolet}\href{http://www.physics.sk/aps/pubs/2002/aps-2002-52-5-363.pdf}{Acta Physica Slovaca 、\textbf{52} (5) (2002) 363}}, \href{https://arxiv.org/abs/0803.2685}{[arXiv:0803.2685 [gr-qc]]}.

\bibitem{0307049}
Zden\v{e}k Stuchl\'ik,  Petr Slan\'y, ``Equatorial Circular Orbits in the Kerr-de Sitter Spacetimes'', {\changeurlcolor{vividviolet}\href{http://journals.aps.org/prd/abstract/10.1103/PhysRevD.69.064001}{Phys. Rev. D \textbf{69} (2004) 064001}}, \href{https://arxiv.org/abs/gr-qc/0307049}{[arXiv:gr-qc/0307049]}.

\bibitem{0511057}
Andres Balaguera-Antolinez, Christian G. Boehmer, Marek Nowakowski, ``Scales Set by the Cosmological Constant'', {\changeurlcolor{vividviolet}\href{http://iopscience.iop.org/article/10.1088/0264-9381/23/2/013/meta;jsessionid=8DA7B6EEDCF730F3B14CB5F347F60EC8.c1.iopscience.cld.iop.org}{Class. Quant. Grav. \textbf{23} (2006) 485}}, \href{https://arxiv.org/abs/gr-qc/0511057}{[arXiv:gr-qc/0511057]}.

\bibitem{1005.1107}
Ivan Arraut, Davide Batic, Marek Nowakowski, ``Velocity and Velocity Bounds in Static Spherically Symmetric Metrics'', {\changeurlcolor{vividviolet}\href{https://www.degruyter.com/view/j/phys.2011.9.issue-4/s11534-010-0147-0/s11534-010-0147-0.xml}{Central Eur. J. Phys. \textbf{9}  (2011) 926}}, \href{https://arxiv.org/abs/1005.1107}{[arXiv:1005.1107 [gr-qc]]}.



\bibitem{1608.02202}
Mirjam Cvetic, Gary W. Gibbons, Christopher N. Pope, ``Photon Spheres and Sonic Horizons in Black Holes from Supergravity and Other Theories'', \href{https://arxiv.org/abs/1608.02202}{[arXiv:1608.02202 [gr-qc]]}.


\bibitem{claudel}
Clarissa-Marie Claudel, Kumar S. Virbhadra, George F.R. Ellis, ``The Geometry of Photon Surfaces'', {\changeurlcolor{vividviolet}\href{http://scitation.aip.org/content/aip/journal/jmp/42/2/10.1063/1.1308507}{J. Math. Phys. \textbf{42} (2001) 818}}, \href{http://arxiv.org/abs/gr-qc/0005050}{[arXiv:gr-qc/0005050]}.

\bibitem{calvani} Massimo Calvani, Fernando de Felice,``Vortical Null Orbits, Repulsive Barriers, Energy Confinement in Kerr Metric'', {\changeurlcolor{vividviolet}\href{http://link.springer.com/article/10.1007\%2FBF00759648}{Gen. Rel. Grav. \textbf{9} (1978) 889}}.

\bibitem{stuchlik}
Zden\v ek Stuchl\'ik, ``The Radial Motion of Photons in Kerr Metric'', {\changeurlcolor{vividviolet}\href{http://adsabs.harvard.edu/full/1981BAICz..32...40S}{Bull. Astronom. Inst. Czechoslovakia \textbf{32} (1981) 40}}.

\bibitem{calvani2}
Massimo Calvani, Fernando de Felice, Luciano Nobili, ``Photon Trajectories in the Kerr-Newman Metric'', {\changeurlcolor{vividviolet}\href{https://iopscience.iop.org/article/10.1088/0305-4470/13/10/018/pdf}{J. Phys. A: Math. Gen. \textbf{13} (1980) 3213}}.

\bibitem{SimRose}
Michael Simpson, Roger Penrose, ``Internal Instability in a Reissner-Nordstr\"om Black Hole'', {\changeurlcolor{vividviolet}\href{http://link.springer.com/article/10.1007\%2FBF00792069}{Int. J. Theor. Phys. \textbf{7} (1973) 183}}.

\bibitem{poisson}
Eric Poisson, Werner Israel, ``Inner-Horizon Instability and Mass Inflation in Black Holes'', {\changeurlcolor{vividviolet}\href{http://journals.aps.org/prl/abstract/10.1103/PhysRevLett.63.1663}{Phys. Rev. Lett. \textbf{63} (1989) 1663}}.

\bibitem{1501.04598}
Jonathan Luk, Sung-Jin Oh, ``Proof of Linear Instability of the Reissner-Nordstr\"{o}m Cauchy Horizon Under Scalar Perturbations'', \href{http://arxiv.org/abs/1501.04598}{[arXiv:1501.04598 [gr-qc]]}.



\bibitem{1107.5081}
Ted Jacobson, ``Where is the Extremal Kerr ISCO?'', {\changeurlcolor{vividviolet}\href{https://iopscience.iop.org/article/10.1088/0264-9381/28/18/187001/meta}{Class. Quant. Grav. \textbf{28} (2011) 187001}}, \href{http://arxiv.org/abs/1107.5081}{[arXiv:1107.5081 [gr-qc]]}. 

\bibitem{1503.01973}
Sebastian Ulbricht, Reinhard Meinel, ``Note on Circular Geodesics in the Equatorial Plane of an Extreme Kerr-Newman Black Hole'', {\changeurlcolor{vividviolet}\href{https://iopscience.iop.org/article/10.1088/0264-9381/32/14/147001/meta}{Class. Quant. Grav. \textbf{32} (2015) 14, 147001}}, \href{http://arxiv.org/abs/1503.01973}{[arXiv:1503.01973 [gr-qc]]}.

\bibitem{doran}
Chris Doran, ``A New Form of the Kerr Solution'', {\changeurlcolor{vividviolet}\href{http://journals.aps.org/prd/abstract/10.1103/PhysRevD.61.067503}{Phys. Rev. D \textbf{61} (2000) 067503}}, \href{http://arxiv.org/abs/gr-qc/9910099}{[arXiv:gr-qc/9910099]}.

\bibitem{9709224}
Raphael Bousso, Stephen Hawking, ``(Anti-)Evaporation of Schwarzschild-de Sitter Black Holes'', {\changeurlcolor{vividviolet}\href{https://journals.aps.org/prd/abstract/10.1103/PhysRevD.57.2436}{Phys. Rev. D \textbf{57} (1998) 2436}}, \href{https://arxiv.org/abs/hep-th/9709224}{[arXiv:hep-th/9709224]}.

\bibitem{0712.3315v2}
Yun Soo Myung, ``Thermodynamics of Schwarzschild-de Sitter Black Hole: Thermal Stability of Nariai Black Hole'', {\changeurlcolor{vividviolet}\href{https://journals.aps.org/prd/abstract/10.1103/PhysRevD.77.104007}{Phys. Rev. D \textbf{77} (2008) 104007}}, \href{https://arxiv.org/abs/0712.3315v2}{[arXiv:0712.3315 [gr-qc]]}.

\bibitem{1502.02737}
Sean Stotyn, ``A Tale of Two Horizons'', {\changeurlcolor{vividviolet}\href{http://www.nrcresearchpress.com/doi/10.1139/cjp-2015-0091#.V9juJvl95D8}{Can. J. Phys. \textbf{93} (2015) 995}}, \href{https://arxiv.org/abs/1502.02737}{[arXiv:1502.02737 [gr-qc]]}. 


\bibitem{livrev}
Geoffrey Comp\`{e}re, 
``The Kerr/CFT Correspondence and its Extensions'', 
{\changeurlcolor{vividviolet}\href{http://relativity.livingreviews.org/open?pubNo=lrr-2012-11&amp;page=title.html}{Living Rev. Relativity \textbf{15} (2012) 11}}, \href{http://arxiv.org/abs/1203.3561}{[arXiv:1203.3561 [hep-th]]}. 

\bibitem{gttgrr}
Ted Jacobson, ``When is $g_{tt} g_{rr} = -1$?'', {\changeurlcolor{vividviolet}\href{https://iopscience.iop.org/article/10.1088/0264-9381/24/22/N02/meta}{Class. Quant. Grav. \textbf{24} (2007) 5717}}, \href{http://arxiv.org/abs/0707.3222v3}{[arXiv:0707.3222 [gr-qc]]}.

\bibitem{HCR1}
Brandon Carter, ``Axisymmetric Black Hole Has Only Two Degrees of Freedom'', {\changeurlcolor{vividviolet}\href{http://journals.aps.org/prl/abstract/10.1103/PhysRevLett.26.331}{Phys. Rev. Lett. \textbf{26} (1971) 331}}.

\bibitem{HCR2}
Stephen W. Hawking, George F.R. Ellis, \emph{The Large Scale Structure of Space-Time}, Cambridge Univ. Press, 1973.

\bibitem{HCR3}
David C. Robinson, ``Uniqueness of the Kerr Black Hole'', {\changeurlcolor{vividviolet}\href{http://journals.aps.org/prl/abstract/10.1103/PhysRevLett.34.905}{Phys. Rev. Lett. \textbf{34} (1975) 905}}.

\bibitem{K1}
Alexandru D. Ionescu, Sergiu Klainerman, ``On the Uniqueness of Smooth, Stationary Black Holes in Vacuum'', {\changeurlcolor{vividviolet}\href{http://link.springer.com/article/10.1007\%2Fs00222-008-0146-6}{Invent. Math. \textbf{175} (2008) 35}}, \href{http://arxiv.org/abs/0711.0040}{[arXiv:0711.0040 [gr-qc]]}.

\bibitem{K2}
Spyros Alexakis, Alexandru D. Ionescu, Sergiu Klainerman, ``Uniqueness of Smooth Stationary Black Holes in Vacuum: Small Perturbations of the Kerr Spaces'', {\changeurlcolor{vividviolet}\href{http://link.springer.com/article/10.1007\%2Fs00220-010-1072-1}{Commun. Math. Phys. \textbf{299} (2010) 89}}, \href{http://arxiv.org/abs/0904.0982}{[arXiv:0904.0982 [gr-qc]]}.


\bibitem{zum}
Henning M\"uller zum Hagen, ``On the Analyticity of Static Vacuum Solutions of Einstein's Equations'', {\changeurlcolor{vividviolet}\href{http://journals.cambridge.org/action/displayAbstract?fromPage=online&aid=2069152&fulltextType=RA&fileId=S0305004100045710}{Math. Proc. Camb. Phil. Soc. \textbf{67} (1970) 415}}.


\bibitem{0402087}
Piotr T. Chru\'sciel, ``On Analyticity of Static Vacuum Metrics at Non-Degenerate Horizons'', {\changeurlcolor{vividviolet}\href{http://www.actaphys.uj.edu.pl/fulltext?series=Reg&vol=36&page=17}{Acta Phys. Polon. B \textbf{36} (2005) 17}}, \href{https://arxiv.org/abs/gr-qc/0402087v2}{[arXiv:gr-qc/0402087]}.


\bibitem{0408016v1}
Norman Cruz, Marco Olivares, Jose R. Villanueva, 
``The Geodesic Structure of the Schwarzschild Anti-de Sitter Black Hole'', {\changeurlcolor{vividviolet}\href{https://iopscience.iop.org/article/10.1088/0264-9381/22/6/016/meta}{ Class. Quant. Grav. \textbf{22} (2005) 1167}}, \href{https://arxiv.org/abs/gr-qc/0408016v1}{[arXiv:gr-qc/0408016]}.

\bibitem{GHS}
David Garfinkle, Gary T. Horowitz, Andrew Strominger, ``Charged Black Holes in String Theory'', {\changeurlcolor{vividviolet}\href{http://journals.aps.org/prd/abstract/10.1103/PhysRevD.43.3140}{Phys. Rev. D \textbf{43} (1991) 31403143}}.

\bibitem{GM}
Gary W. Gibbons, Kei-ichi Maeda, ``Black Holes and Membranes in Higher Dimensional Theories with Dilaton Fields'', {\changeurlcolor{vividviolet}\href{http://www.sciencedirect.com/science/article/pii/0550321388900065}{Nucl. Phys. B \textbf{298} (1981) 741}}.

\bibitem{darkside}
Gary T. Horowitz, ``The Dark Side of String Theory: Black Holes and Black Strings'', Proceedings of String Theory and Quantum Gravity '92, p.55-99, \href{http://arxiv.org/abs/hep-th/9210119}{[arXiv:hep-th/9210119]}.

\bibitem{1109.0254}
Sharmanthie Fernando, ``Null Geodesics of Charged Black Holes in String Theory'', {\changeurlcolor{vividviolet}\href{http://journals.aps.org/prd/abstract/10.1103/PhysRevD.85.024033}{Phys. Rev. D \textbf{85} (2012) 024033}}, \href{http://arxiv.org/abs/1109.0254}{[arXiv:1109.0254 [hep-th]]}.

\bibitem{9202014}
Christoph F.E. Holzhey, Frank Wilczek, ``Black Holes as Elementary Particles '', {\changeurlcolor{vividviolet}\href{http://www.sciencedirect.com/science/article/pii/0550321392902549}{Nucl. Phys. B \textbf{380} (1992) 447}}, \href{http://arxiv.org/abs/hep-th/9202014v1}{[arXiv:hep-th/9202014]}.


\bibitem{O'Neill}
Barrett O'Neill, \emph{The Geometry of Kerr Black Holes}, A K Peters, Ltd., 1995.

\bibitem{hobson}
Michael Hobson, George Efstathiou, Anthony Lasenby, \emph{General Relativity: An Introduction for Physicists}, 1st Edition, Cambridge University Press, 2006.

\bibitem{DK}
Naresh Dadhich, P. P. Kale, ``Equatorial Circular Geodesics in the Kerr-Newman Geometry'', {\changeurlcolor{vividviolet}\href{http://adsabs.harvard.edu/full/1981BAICz..32..366S}{J. Math. Phys. \textbf{18} (1977) 1727}}.

\bibitem{S1981}
Zden\v{e}k Stuchl\'ik, ``Null Geodesics in the Kerr-Newman Metric'',  {\changeurlcolor{vividviolet}\href{http://adsabs.harvard.edu/full/1981BAICz..32..366S}{Bull. Astronom. Inst. Czechoslovakia \textbf{32} (1981) 366}}.

\bibitem{BBS}
Vladimir Balek, Ji\v{r}\'i Bi\v{c}\'ak, Zden\v{e}k Stuchl\'ik, ``The Motion of the Charged Particles in the Field of Rotating Charged Black Holes and Naked Singularities. II - The Motion in the Equatorial Plane'', {\changeurlcolor{vividviolet}\href{http://adsabs.harvard.edu/full/1989BAICz..40..133B}{Bull. Astronom. Inst. Czechoslovakia \textbf{40} (1989) 133}}.

\bibitem{1406.1295}
Sarani Chakraborty, Asoke  K. Sen, ``Light Deflection Due to a Charged, Rotating Body'', {\changeurlcolor{vividviolet}\href{https://iopscience.iop.org/article/10.1088/0264-9381/32/11/115011/meta}{Class. Quant. Grav. \textbf{32} (2015) 115011}}, \href{https://arxiv.org/abs/1406.1295}{[arXiv:1406.1295 [gr-qc]]}.



\bibitem{khrip}
Iosif B. Khriplovich, ``Nonthermal Radiation from Black Holes'', {\changeurlcolor{vividviolet}\href{http://link.springer.com/article/10.1134\%2F1.1495025}{Phys. Atom. Nucl. \textbf{65} (2002) 1259}}.

\bibitem{1206.0120}
Brett McInnes, ``Universality of the Holographic Angular Momentum Cutoff'', {\changeurlcolor{vividviolet}\href{http://www.sciencedirect.com/science/article/pii/S0550321312003987}{Nucl. Phys. B \textbf{864} (2012) 722}}, \href{http://arxiv.org/abs/1206.0120}{[arXiv:1206.0120 [hep-th]]}.

\bibitem{1403.3258}
Brett McInnes, ``Angular Momentum in QGP Holography'', {\changeurlcolor{vividviolet}\href{http://www.sciencedirect.com/science/article/pii/S0550321314002661}{Nucl. Phys. B \textbf{887} (2014) 246}}, \href{http://arxiv.org/abs/1403.3258}{[arXiv:1403.3258 [hep-th]]}.

\bibitem{AB}
Lars Andersson, Pieter Blue, ``Hidden Symmetries and Decay for the Wave
Equation on the Kerr Spacetime'', {\changeurlcolor{vividviolet}\href{http://annals.math.princeton.edu/2015/182-3/p01}{Annals of Mathematics \textbf{182} (2015) 787}}, \href{http://arxiv.org/abs/0908.2265}{[arXiv:0908.2265 [math.AP]]}.

\bibitem{1103.2355}
Irene Bredberg, Cynthia Keeler, Vyacheslav Lysov, Andrew Strominger, ``Cargese Lectures on the Kerr/CFT Correspondence'', {\changeurlcolor{vividviolet}\href{http://www.sciencedirect.com/science/article/pii/S0920563211004312}{Nucl. Phys. Proc. Suppl. \textbf{216} (2011) 194}}, \href{http://arxiv.org/abs/1103.2355v3}{[arXiv:1103.2355 [hep-th]]}.

\bibitem{4500311}
Shanjit Heisnam, Irom Ablu Meitei,  Kangujam Yugindro Singh, ``Motion of a Test Particle in the Kerr-Newman De/Anti De Sitter Space-Time'', {\changeurlcolor{vividviolet}\href{http://file.scirp.org/Html/6-4500311_46541.htm}{IJAA \textbf{4} (2014) 365}}.

\bibitem{1009.1661}
Bradly K. Button, Leo Rodriguez, Catherine A. Whiting, Tuna Yildirim, ``A Near Horizon CFT Dual for Kerr-Newman-AdS'', 	{\changeurlcolor{vividviolet}\href{http://www.worldscientific.com/doi/abs/10.1142/S0217751X11053663}{Int. J. Mod. Phys. A \textbf{26} (2011) 3077}}, \href{http://arxiv.org/abs/1009.1661}{[arXiv:1009.1661 [hep-th]]}.

\end{thebibliography}
\end{document}